\documentclass[11pt] {article}
\usepackage[utf8]{inputenc}
\usepackage{amsmath}

\usepackage[margin=1.25in]{geometry}

\usepackage{setspace}
\usepackage[semicolon]{natbib}
\usepackage[all]{xy}
\usepackage{color}
\xyoption{color}

\usepackage{tikz-network}

\usepackage{caption}
\usepackage{dirtytalk}

\usepackage{color}
\usepackage[colorlinks=true,allcolors=blue,backref=page, hyperindex,breaklinks]{hyperref}
\usepackage{amsmath, amssymb, amsthm}
\usepackage{mathrsfs}
\usepackage{mathtools}
\usepackage[noabbrev,capitalize,nameinlink]{cleveref}
\crefname{equation}{}{}
\usepackage{fullpage}
\usepackage{graphics}
\usepackage{pifont}
\usepackage{tikz}
\usepackage{bbm}
\usepackage[T1]{fontenc}
\DeclareMathOperator*{\argmax}{arg\,max}

\usetikzlibrary{arrows.meta}

\usepackage{environ}
\usepackage{framed}
\usepackage{url}
\usepackage[linesnumbered,ruled,vlined]{algorithm2e}
\usepackage[noend]{algpseudocode}
\usepackage[labelfont=bf]{caption}
\usepackage{framed}
\usepackage[framemethod=tikz]{mdframed}
\usepackage{appendix}
\usepackage{graphicx}
\usepackage[textsize=tiny]{todonotes}
\usepackage[most]{tcolorbox}
\usepackage{enumerate}
\usepackage{verbatim}
\usepackage{algpseudocode}
\allowdisplaybreaks[1]

\apptocmd{\sloppy}{\hbadness 10000\relax}{}{} 

\crefname{algorithm}{Algorithm}{Algorithms}

\crefname{algocf}{Algorithm}{Algorithms}

\crefname{equation}{}{} 
\crefname{conjecture}{Conjecture}{Conjectures} 
\AtBeginEnvironment{appendices}{\crefalias{section}{appendix}} 

\usepackage[color,final]{showkeys} 

\colorlet{refkey}{orange!20}
\colorlet{labelkey}{blue!30}

\numberwithin{equation}{section}
\newtheorem{theorem}{Theorem}[section]
\newtheorem{proposition}[theorem]{Proposition}
\newtheorem{lemma}[theorem]{Lemma}

\crefname{claim}{Claim}{Claims}

\newtheorem{corollary}[theorem]{Corollary}

\newtheorem*{question*}{Question}
\newtheorem{fact}[theorem]{Fact}

\theoremstyle{definition}
\newtheorem{definition}[theorem]{Definition}

\newtheorem*{definition*}{Definition}
\newtheorem{example}[theorem]{Example}

\theoremstyle{remark}

\newcommand{\abs}[1]{\left\lvert#1\right\rvert}

\newcommand{\floor}[1]{\left\lfloor #1 \right\rfloor}
\newcommand{\ceil}[1]{\left\lceil #1 \right\rceil}

\newcommand{\CC}{\mathcal{C}}
\newcommand{\C}{\mathcal{C}}
\newcommand{\B}{\mathcal{B}}
\newcommand{\E}{\mathcal{E}}
\newcommand{\M}{\mathcal{M}}
\newcommand{\I}{\mathcal{I}}
\newcommand{\V}{\mathcal{V}}
\newcommand{\PP}{\mathcal{P}}

\allowdisplaybreaks

\SetKwInput{KwInput}{Input}                
\SetKwInput{KwOutput}{Output}              

\usepackage{ifthen}
\newboolean{demand_law_on}
\setboolean{demand_law_on}{false}

\vspace{-2cm}
\title{Reserve Matching with Thresholds}
\author{Suat Evren\footnote{Departments of Mathematics, Computer Science, and Economics, Massachusetts Institute of Technology (email: evrenis@mit.edu). 
 I am truly indebted to Yannai Gonczarowski and Scott Kominers for advising and fruitful discussions that significantly contributed to the paper. I am also grateful to Henry Cohn, Yunus Semih Co\c{s}kun, O\u{g}uzhan \c{C}elebi, Federico Echenique, S\"{u}leyman Kerimov, Joanna Kondylis, Youssef Marrakchi, Parag Pathak, Tayfun S\"{o}nmez, Bertan Turhan, Bumin Yenmez, and Muhammed Ali Y{\i}ld{\i}r{\i}m for helpful comments and feedback. I conducted a part of this research when I was a research fellow at Harvard Business School. I gratefully acknowledge financial support from MIT Political Science Department Pressman Award 2022, Harvard Business School PRIMO Fellowship, and MIT UROP direct funding. 
}
\\ MIT}
\date{first draft: May 2023, this version: April 2025} 

\begin{document}

\maketitle
\begin{abstract}
We develop a general framework for reserve systems that allocate scarce resources such as vaccines to unit-demand agents under prioritization and eligibility constraints, along with a computationally efficient mechanism. Reserve systems allocate scarce resources --such as vaccines, medical units, school seats, or government positions-- to essential groups by creating categories with prioritized beneficiaries. Prior work typically assumed a common baseline priority ordering and featured either hard or soft reserves. The threshold reserve model we introduce supports independent priority orderings, mixtures of hard and soft reserves, and overlapping categories, thereby capturing both beneficiary designations and eligibility constraints while offering policymakers greater flexibility. Our Iterative Max-in-Max Assignment Mechanism (IMMAM) satisfies all desirable properties in this domain: it respects priorities within categories, maximizes resource utilization, and then lexicographically maximizes beneficiary assignments. IMMAM is path independent and therefore well-behaved in settings with multiple institutions making simultaneous allocation decisions. We leverage path independence to obtain comparative statics and to significantly improve the mechanism's computational efficiency. We outline applications of our framework in the context of vaccine allocation.
\end{abstract}



\section{Introduction}\label{intro}
Many important problems in economics consist of reallocation of indivisible resources such as distributing vaccines in a pandemic, or assigning minority students to elite colleges in India. When distributing scarce resources, economists commonly aim to fulfill some fairness and optimality constraints in the eventual allocation. Traditionally, Pareto optimality has been the primary optimality constraint used. However, in recent literature on matching theory and reserve systems, a novel class of optimization based mechanisms called \say{smart matching mechanisms} have been introduced.\footnote{The term \say{smart markets} in the field of management science has conventionally referred to complex and information-intensive markets that rely on computational optimization of an objective function. Its earliest known usage can be traced to a sequence of seminal papers, which propounded a novel auction design for natural gas markets subsequent to the emerging need upon the enactment of the Natural Gas Policy Act \citep*{McCabe...89, McCabe...91}. 
Subsequently, \cite{Glazer99} used it to characterize markets that are both dynamic and information-intensive. Notably, Glazer also denominated the back-then traditional markets as \say{dumb markets,} which he characterizes as \say{static, fixed, and basically information-poor.} In a more recent research commentary, \cite*{BGK10} drew attention to the importance of \say{theoretically supported computational tools} and algorithmic mechanism design in smart markets. It is worth noting that smart market solutions have tended to rely predominantly on linear programming approaches until the last decade. In the modern field of market design, \cite{Roth07} used the term \say{smart markets} in his review article \emph{The Art of Designing Markets}. He used it to describe markets that \say{combine the inputs of users in complex ways} by using computers, giving the example of kidney exchange \citep*{kidney04, kidney07} since \say{by running through every possible combination of donors and patients, [the mechanism] can arrange the highest possible number of transplants.} Recently, \say{smart market design} has surfaced in the market design literature, offering novel mechanisms in the context of smart reserve systems. The earliest of these mechanisms was the \emph{Meritorious Horizontal Choice Rule} initially proposed in the domain of Indian affirmative action and school choice \citep{SY22}. Subsequently, Smart Reserve Matching was introduced in the context of medical rationing and vaccine allocation \citep{PSUY20}, which constitutes a small modification on the former mechanism. To the best of our knowledge, our paper constitutes the third documented instance of a smart reserve system. It is also the first smart reserve mechanism that does not assume baseline priority orderings.}

Reserve systems are used to allocate scarce resources to individuals while ensuring that no segment of the population is under-represented in the final allocation. Some examples include allocating school seats at elite high schools and colleges for minority students \citep*{Hafalir...13, AB21, AT23}, allocating homogeneous government positions to a set of workers with diverse backgrounds \citep*{SY22}, rationing medical resources while prioritizing different categories of essential workers such as healthcare professionals and national security workers at the same time \citep*{PSUY20}, or satisfying tailored diversity requirements in public programs \citep*{gonczarowski2020matching}. For instance, during the COVID-19 pandemic, there was a discussion of whom to prioritize first for vaccination. By using a reserve system, we can create different categories for healthcare professionals, the elderly, and the infants separately, by which we can accommodate different essential segments of the population simultaneously. Each category gets its name from the group of beneficiaries it aims to serve. 

In the classic model of a reserve system,\footnote{Reserve systems were first introduced in \cite{Hafalir...13}. A general theory of the classic reserve systems framework was developed in \cite{PSUY20}.} agents have unit demand, and each category is reserved a certain number of identical units. While agents are indifferent between categories, each category has a priority ordering of agents. Besides, each category has a set of \emph{beneficiaries}, which corresponds to a top segment of the priority ordering in the respective category. A \emph{hard} reserve system is a system under which an agent must be a beneficiary of a category to get a unit from that category. A \emph{soft} reserve system is a system under which a non-beneficiary can receive a unit if there is excess supply of reserves in a category.\footnote{For a discussion on hard versus soft reserves, see \cite*{EHYY14}.} When allocating vaccines, it might be better to use a soft reserve system since we would not like to waste any vaccine.

The classic model of reserve systems, as we described above, effectively creates two tiers of statues for agents in each category. However, the two-tier solution might not be sufficient in certain cases. For example, in the vaccine allocation problem, the specific vaccine we use for the elderly category may not be safe for the pediatric patients whereas it is safe for the adults, in which case we need three tiers: beneficiaries (the elderly), eligible (adults), and non-eligible (infants).\footnote{We discuss the problem of vaccine allocation in \cref{application: vaccine} in more detail.} Another real-life example is the Indian affirmative action procedure that mandates admissions to technical colleges in India where it is required by law for some categories to use soft reserves and others to use hard reserves \citep{BusinessRules23}. As a result, we propose the adaptation of the \emph{threshold reserve model}, which offers a more comprehensive three-tier solution to the reserve problem and allows using soft and hard reserves simultaneously.

A reserve system is called \emph{smart} when the designer aims to choose the matching that maximizes the number of beneficiaries receiving vaccines while obeying some fairness constraints. Previous work on smart reserve systems assumed \emph{baseline priorities}. A baseline priority is a master priority ordering, which  can be determined by an exam score in the case of a school choice market \citep{SY22}, or can be a random lottery in the case of a medical unit allocation \citep{PSUY20}. For any category, the beneficiaries are placed above the non-beneficiaries, but they follow this baseline priority ordering within themselves. The non-beneficiaries also preserve the same order among each other. Although assuming a baseline priority ordering through lotteries may not be desirable, obtaining a baseline priority ordering without using lotteries is not feasible in some cases. For example, in the case of medical unit allocation, infants and adults are virtually incomparable since the indicators that are used to measure health of adults are not reliable when used on infants, according to Massachusetts guidelines (pages 17-20, \citeyear{MassGuideline20}).

As a solution, we propose a new smart reserve system for allocating scarce resources that aims to simultaneously optimize over the objectives of maximizing the number of units allocated and the number of beneficiary assignments. Our mechanism achieves all previously proposed desirable properties (and more) in a more general domain than the previously proposed reserve systems, allowing different categories in the reserve system to have completely independent priority orderings in the model we develop. We call this property by the generic name of \emph{universal domain} desideratum, which we achieve by the virtue of the novel \emph{Iterative Max-in-Max Mechanism (IMMAM)}.

While our mechanism can be applied to any resource allocation market, we focus our attention on the vaccine allocation market. Contextualizing our results, we adapt the language of the vaccine allocation problem, closely aligned with the language developed by \cite{PSUY20}, to be consistent with the recent reserve systems literature. We also think that following this language rather than using a more abstract one makes it easier to convey the general idea of our paper. In \cref{application: vaccine}, we give more context regarding the relevance of our model to the problem of vaccine allocation.

In \cref{model}, we formally introduce the threshold reserve model, which constitutes an extension of the reserve system framework. In our approach, we do not assume homogeneity of vaccines, but we do assume that patients exhibit indifference towards different vaccines as long as the vaccine is safe for them. For each category, we establish beneficiary and eligibility thresholds, assign a specific number of vaccines, and establish a priority ordering. The priority orderings and thresholds are permitted to vary arbitrarily across categories. We assume that, for each category, the same type of vaccine is used and that patients who meet the eligibility threshold for that vaccine are safely able to receive it. The role of beneficiary thresholds is more nuanced. Patients who surpass the beneficiary threshold for a particular category are labeled as beneficiaries of that category. These individuals are the primary target of the respective categories. We aim to maximize the number of beneficiaries that are matched with available vaccines, which we term as \emph{maximum in beneficiary assignment}, consistent with prior literature on smart reserve systems. In addition, we introduce another desideratum, \emph{maximum in resource allocation}, which seeks to formalize the notion of allocating as many vaccines as possible to eligible patients, regardless of whether they are beneficiaries or not.

 We formally introduce properties that are desirable in the final allocations of vaccines in \cref{axioms}, which we categorize into two groups as \emph{feasibility axioms} and \emph{desiderata}. The feasibility axioms require a matching to comply with eligibility restrictions and respect priorities within categories. We treat these axioms as hard constraints throughout the paper. The three desiderata consist of the aforementioned universal domain property together with the two maximum conditions.

We then turn to our characterization results in \cref{characterizations}. First, we demonstrate an impossibility result which reveals that no mechanism can simultaneously satisfy all of the desiderata. This finding indicates that if a market designer intends to create a mechanism that is broadly applicable while vaccinating the maximum number of patients feasible, it may be necessary to compromise on the fundamental objective of maximizing the number of beneficiaries receiving vaccines from their respective categories.

\ifthenelse{\boolean{demand_law_on}}{

Next, we show that, under reasonable conditions that we call \emph{demand law}, it is possible to allocate all available vaccines to beneficiaries. Moreover, IMMAM is guaranteed to produce such an allocation when the conditions for the demand law are satisfied. The demand law demonstrates that by narrowing the domain and forsaking the universal domain desideratum, it is possible to satisfy the two optimality criteria. This finding is particularly relevant since most real-world situations meet these conditions, and it is often possible to identify a matching that assigns every unit to a beneficiary. The conditions under which this result holds are especially pertinent when supply is limited but demand remains substantial across all population segments, which is a natural and realistic assumption given the scarcity of resources in the allocation problem. To illustrate this point, we provide an example of a pandemic-related design in the United States in \cref{appendix: on demand law}, where we also show that the sufficient conditions imposed by the demand law are, although not if-and-only-if, tight. 

}{}

As an aid to overcome the impossibility result while preserving the generality of the domain, we propose replacing the two optimality desiderata with a new \emph{max-in-max} property. This property asks the market designer to select an allocation that maximizes the number of beneficiary assignments among the outcomes that achieve the maximum possible resource allocation. In \cref{Max-in-Max Subsection}, we present an algorithm to find a max-in-max matching. However, this algorithm does not necessarily output an outcome that respects priorities.

In \cref{section: IMMAM}, we present the Iterative Max-in-Max Assignment Mechanism (IMMAM), which also respects priorities while outputting a max-in-max matching. It also permits the most general input possible without imposing interdependencies between distinct categories, thereby satisfying the universal domain desideratum. IMMAM iterates over patient-category pairs by first fixing a category according to an arbitrary precedence list provided as input, and then traversing the patients in that category's priority list. For each patient-category pair, it checks whether there exists a max-in-max matching—respecting all previously made permanent assignments—in which the patient is assigned a vaccine from the given category. If such a matching exists, the mechanism permanently assigns the patient to the category, reduces the category's reserve by one, and moves on to the next pair in the sequence.

The set of patients who are assigned a vaccine in the outcome of IMMAM defines a choice function. We show that this choice function satisfies \emph{path independence}. In the presence of multiple institutions that allocate vaccines and where patients have preferences over the institutions, path independence implies that there exists a stable outcome \citep{CY17}. These institutions can be different brands of vaccines, different hospitals, or pharmacies. In addition, path independence has normatively desirable comparative statics implications. In particular, it implies that if a patient leaves the market, every other patient who was assigned a vaccine still receives a vaccine. Moreover, we show that the same holds if the reserves in any one of the categories increase, possibly due to a boost in vaccine production.

In \cref{speeding up the algorithm}, we improve the runtime complexity of our mechanism in two steps. The first optimization we introduce gets rid of the redundancies caused by duplicating some of the work that the algorithm has already done. We call this optimized version \emph{IMMAM with Memory (IMMAM-M)}. The second optimization technique we use signifies the algorithmic implications of path independence. In particular, since the number of patients assigned a vaccine is bounded by the total number of available vaccines, we can partition the set of patients into smaller subsets, apply IMMAM-M to each subset, and recursively run IMMAM-M by gradually combining the resulting sets of chosen patients. While doing so, we can still ensure that the final set of patients who are assigned a vaccine is the same as the one in IMMAM thanks to path independence. We call this further optimized version \emph{IMMAM-M with Bootstrapping (IMMAM-MB)}. In the end, we obtain a runtime of $O(q \cdot |\PP| \cdot |\CC|^4)$, where $q$ is the total number of vaccines, and $\PP$ and $\CC$ denote the sets of patients and categories, respectively. Note that in all practical applications of reserve systems, the number of categories is only a handful regardless of the amount of resources or individuals in the market. Thus, the runtime can arguably considered to be $O(q \cdot |\PP|)$.

In \cref{section: review}, we review the literature. First, we discuss the related theoretical literature in \cref{subsection: related literature}. Then, we provide more contextual details and background on the problem of vaccine allocation in \cref{application: vaccine}. Readers who would like to gain more insight in preparation for the development of our model in \cref{model} are encouraged to read \cref{application: vaccine} first.

In \cref{conclusion}, we conclude. All of the proofs are relegated to \cref{appendix: proofs}. Omitted algorithms are given in \cref{appendix: omitted algorithms}. \ifthenelse{\boolean{demand_law_on}}{Some further discussion on the demand law can be found in \cref{appendix: on demand law}.}{}

\section{The Threshold Reserve Model} \label{model}

In describing our model, we will use the language of vaccine allocation. 
In a vaccine allocation market, there is a set of \textbf{patients} $\PP$ and $q$ \textbf{vaccines} to allocate. There is a set of \textbf{reserve categories} $\CC$. For every category $c \in \CC$, $r_c$ units are \textbf{reserved} for that category so that $\sum_{c \in \CC} r_c = q$.\footnote{Note that this is without loss of generality since we can always create a general category that has everybody as a beneficiary and reserve the remaining vaccines for that category.}

For every category $c \in \CC$, there is a \textbf{beneficiary threshold} $\beta$ and an \textbf{eligibility threshold} $\eta$ with a strict \textbf{priority ordering} $\pi_c$ over the disjoint union $\PP \cup \{ \beta, \eta \}$ such that $\beta \: \pi_c \: \eta$. For every patient $p \in \PP$, we say that $p$ is a \textbf{beneficiary} of category $c$ if
    $p \: \pi_c \: \beta.$
Similarly, we say that $p$ is $\textbf{eligible}$ for category $c$ if
    $p \: \pi_c \: \eta.$
For any category $c \in \CC$, we denote the set of beneficiaries by $\B_c$, and the set of eligible patients by $\E_c$. 

It is worth emphasizing that the model admits any arbitrary $\pi_c$ for each $c \in \CC$ as long as the condition
    $\beta \: \pi_c \: \eta$
is satisfied. An equivalent way of writing this condition is that $\B_c \subseteq \E_c$. It is to say that if a patient is a beneficiary of category $c$, then she must be eligible for this category in the first place. 

It is important to discuss the distinction between eligibility and beneficiary and why they could be different. In particular, the set of eligible patients for a category represents the patients who can safely receive this vaccine. For example, some vaccines might not be FDA approved for certain age groups. As was the case with the Pfizer-BioNTech COVID-19 vaccine, it was FDA approved for emergency use in the 12-15 age group as of May 2021, while the Moderna vaccine was authorized only for individuals aged 17 and up \citep{Pfizer-for-adolescents}. Under this scenario, we might want to reserve the Moderna vaccines for a category in which those under 17 are not eligible. On the other hand, beneficiaries of a category are the patients we want to emphasize the most in a category. We try to assign patients vaccines from categories of which they are beneficiaries; however, we also retain the option to assign a vaccine to a patient who is not a beneficiary but still can safely receive this vaccine in case we have excess vaccines. This approach effectively combines the notion of hard reserve systems and soft reserve systems under one umbrella, and lets us find unified and intermediate ways of accounting for the strengths of both reserve types. 

We can consider the vaccine allocation problem in the framework of two-sided matching markets. On one side of the market, we have patients who would like to receive a vaccine; on the other side, we have categories that have priority orderings over patients. In a scenario where receiving a vaccine is crucial, allowing preferences over vaccines may not be worth potentially decreasing the number of matchings in the outcome. Therefore, we do not allow the patients in our model to have preferences over different categories or vaccines. As a result, a \textbf{vaccine allocation market} can be represented by the tuple $(\PP,\C, r,\pi)$, where $r = (r_c)_{c \in \CC}$ and $\pi = (\pi_c)_{c \in \CC}$ succinctly denote the profile of reserves and priorities, respectively.

We define a matching between patients and categories formally as follows. A \textbf{matching} $\mu: \PP \to \C \cup \{\varnothing\}$ is a function that maps each patient either to a category or to $\varnothing$ such that $\abs{\mu^{-1}(c)} \leq r_c$ for every category $c \in \CC$. For any patient $p \in \PP$, we write $\mu(p) = \varnothing$ to mean that the patient does not receive a vaccine and $\mu(p) = c \in \CC$ means that the patient receives a vaccine reserved for category $c$. We denote the set of patients matched under $\mu$ by $\PP_{\mu}$. 
In this model, every patient has unit demand and is indifferent between different categories. 

A \textbf{mechanism} $\M: (\PP, \C, r, \pi) \mapsto  \mu$ is a function that takes a vaccine allocation market as input and outputs a matching $\mu: \PP \to \C \cup \{\varnothing\}.$

\section{Feasibility Axioms and Desiderata} \label{axioms}

There are five main properties that we would seek to satisfy in a mechanism, divided into two categories called \emph{feasibility axioms} and \emph{desiderata}. The two axioms lay out primary conditions for the feasibility of an outcome, whereas the three desiderata are properties that we desire to satisfy in a mechanism. In any possible mechanism we consider, we will take the axioms as a must, but the trade-offs will be observed between the desiderata.

\subsection{Feasibility Axioms} \label{subsection: feasibility axioms}

We start by giving the two axioms with their formal definitions using the notation introduced in \cref{model}, each followed by their interpretations. Note that in all of the results given in this paper, we will treat these three axioms as hard constraints.

\begin{definition}
A matching $\mu$ \textbf{complies with eligibility requirements} if for any $p \in \PP$ and $c \in \C$,
$$ \mu(p) = c \Rightarrow p \in \E_c. $$
\end{definition}

 Our first axiom requires that resources in any category be allocated only to eligible patients. A patient may not be eligible for a category for numerous different reasons. One reason could be that the resources may not be versatile. An infant may not be eligible for the elderly category if the particular vaccine used for the elderly is not appropriate for infants. In the case of labor markets, an individual may not have the necessary skills or training required to perform a job.



\begin{definition}
A matching $\mu$ \textbf{respects priorities} if for any $p, p' \in \PP$ and $c \in \C$,
$$ \mu(p) = c \quad \text{and} \quad \mu(p') = \varnothing \Rightarrow p \: \pi_c \: p' . $$
\end{definition}

The second axiom formulates the idea that if there are two individuals, only one of which has been awarded a unit from a category, the one who has been awarded a unit must have a higher priority than the other.



\subsection{Desiderata} \label{subsection: three desiderata}
Next, we give the three desiderata we would ideally seek to satisfy in a mechanism. However, as we will see in short, it is not possible to satisfy all of them simultaneously. Henceforth, the trade-offs in the results are made between the three desiderata. The first two desiderata relate to the optimality of the outcome, and the last one is regarding the domain of the mechanism.

\begin{definition}\label{maximal in resource allocation}
A matching $\mu$ is \textbf{maximum in resource allocation} if 
$$ \mu \in \argmax_{\mu'} \left| \cup_{c \in \C} \left((\mu')^{-1}(c) \cap \E_c \right)\right|, $$
where $\E_c$ is the set of patients who are eligible for category $c$.
\end{definition}

Our first desideratum formulates the idea that among all feasible matchings, we desire to choose the one that allocates the maximum number of vaccines to patients.\footnote{Maximum in resource allocation imposes a stronger optimality constraint than the widely used \emph{non-wastefulness} requirement in the matching theory literature. Non-wastefulness says that every unit should be utilized so long as there is an individual who has not been rationed a unit yet and is eligible for one of the remaining units. In other words, maximum in resource allocation corresponds to finding a global maximum of the objective function in the domain, whereas non-wastefulness corresponds to finding a local maximum.} 

\begin{definition}
A matching $\mu$ is \textbf{maximum in beneficiary assignment} if 
$$ \mu \in \argmax_{\mu'} \left| \cup_{c \in \C} \left( (\mu')^{-1}(c) \cap \B_c \right) \right|, $$
where $\B_c$ is the set of beneficiaries of category $c$.
\end{definition}

This desideratum is analogous to \cref{maximal in resource allocation}, but instead of optimizing over the number of patients assigned a vaccine, it requires optimizing over the number of beneficiaries assigned a vaccine. Note that when the beneficiary thresholds are exactly the same as the eligibility thresholds -- that is, everybody who is eligible for a category is also a beneficiary of that category -- which corresponds to the domain of \emph{hard reserves}, these two axioms coincide. On the other hand, when there is no eligibility threshold -- that is, everyone is eligible for any category -- the resulting domain corresponds to the domain of \emph{soft reserves}, and the maximum in resource allocation corresponds to assigning all vaccines to patients (or vaccinating everyone in case there are more vaccines than patients). However, in the framework we develop, we consider the most general domain possible, which we call the universal domain.

\begin{definition}\label{universal domain}
A mechanism $\M: (\PP, \C, r, \pi) \mapsto \mu$ satisfies \textbf{universal domain} if its input can have any possible configuration of patients, vaccines, categories, reserves, priorities, and thresholds. That is, $\M$ outputs a matching for any $(\PP, \C, r, \pi)$ such that $0 < |\PP|,|\C| < \infty$, $0< r_c < \infty$ for all $c$, and $\beta \: \pi_c \: \eta$ for all $c$.
\end{definition}

Note that the definitions except for \cref{universal domain} are made for matchings. We can extend these definitions from matchings to mechanisms in the following canonical way: A mechanism $\M$ complies with eligibility requirements, respects priorities, is maximum in resource allocation, and is maximum in beneficiary assignment, if its output always complies with eligibility requirements, respects priorities, is maximum in resource allocation, is maximum in beneficiary assignment, respectively.

\section{An Impossibility Result} \label{characterizations}



We require a mechanism to comply with the feasibility axioms. However, when it comes to the desiderata, they cannot all be satisfied at the same time. We start by giving an illustrative example, which leads to an impossibility result.

\bigskip

\begin{figure}[h]
\centering
\begin{tikzpicture}
    \Vertex[label = $c_1$, x=0, y=2]{A}
    \Vertex[label = $c_2$, x=2, y=2]{B}
    \Vertex[label = $p_1$, x=0, y=0]{C}
    \Vertex[label = $p_2$, x=2, y=0]{D}
    \Edge[color=red, style=dashed](A)(C)
    \Edge[style=dashed](A)(D)
    \Edge[style=dashed](B)(C)

    \Vertex[label = $c_1$, x=5, y=2]{A2}
    \Vertex[label = $c_2$, x=7, y=2]{B2}
    \Vertex[label = $p_1$, x=5, y=0]{C2}
    \Vertex[label = $p_2$, x=7, y=0]{D2}
    \Edge[color=red](A2)(C2)
    \Edge[style=dashed](A2)(D2)
    \Edge[style=dashed](B2)(C2)

    \Vertex[label = $c_1$, x=10, y=2]{A3}
    \Vertex[label = $c_2$, x=12, y=2]{B3}
    \Vertex[label = $p_1$, x=10, y=0]{C3}
    \Vertex[label = $p_2$, x=12, y=0]{D3}
    \Edge[color=red, style=dashed](A3)(C3)
    \Edge(A3)(D3)
    \Edge(B3)(C3)
\end{tikzpicture}
 \caption{Illustration of \cref{impossibility example}. Null matching, maximum in beneficiary assignment matching, and maximum in resource allocation matching are given respectively. Being beneficiary is indicated with a red edge, and eligibility alone is indicated with a black edge.}
\end{figure}

\begin{example} \label{impossibility example}
Suppose there are two categories $c_1$ and $c_2$, and two patients $p_1$ and $p_2$. Suppose $q=2, r_{c_1} = 1, r_{c_2}=1$. Let
\begin{align*}
    \pi_{c_1}: p_1 \: \pi \: \beta \: \pi \: p_2 \: \pi \: \eta 
    \quad \text{and} \quad
    \pi_{c_2}: \beta \: \pi \: p_1 \: \pi \: \eta\: \pi \:p_2.
\end{align*}

Then, the only matching $\mu_{b}$ that is maximum in beneficiary assignment is the matching that assigns patient $p_1$ to a vaccine from category $c_1$; that is, $\mu_{b}(p_1) = c_1$ and $\mu_{b}(p_2) = \varnothing$. On the other hand, the only matching $\mu_{e}$ that is maximum in resource allocation is the matching that assigns patient $p_1$ to a vaccine from category $c_2$, and patient $p_2$ to a vaccine from category $c_1$; that is, $\mu_e(p_1) = c_2$ and $\mu_e(p_2) = c_1$.
Therefore, for the priorities defined above, there is no matching that is both maximum in beneficiary assignment and maximum in resource allocation. \qed
\end{example}

This example leaves us with the following impossibility result. 

\begin{theorem}\label{impossibility result}
There is no mechanism designed in the universal domain that is both maximum in beneficiary assignment and maximum in resource allocation.
\end{theorem}

\ifthenelse{\boolean{demand_law_on}}{
If we want to satisfy maximum in beneficiary assignment and maximum in resource allocation at the same time, the universal domain desideratum is too ambitious. However, the universal domain assumption might not be necessary in many real-life applications. For example, it may be natural to not expect to have a special category that has only a handful of beneficiaries but has many vaccines reserved. Thus, it is desirable to find domains in which we can find mechanisms that satisfy both optimality axioms at the same time. A desirable way to do this is to guarantee the allocation of all vaccines to beneficiaries. 

In \cref{impossibility example}, the main issue preventing us from allocating all of the vaccines to beneficiaries is that one of the categories does not even have any beneficiaries in the first place. In fact, we would face the issue that it is not possible to allocate all of the vaccines to beneficiaries whenever there are many categories that do not have enough beneficiaries, yet too many vaccines are reserved for them. First, we formalize the notion of not having enough beneficiaries, and then give a result that formalizes the idea in the preceding sentence.

\begin{definition}
    We call a category $c$ \textbf{sparse} if it has fewer beneficiaries than the total number of vaccines. That is, $c$ is called sparse if $|\B_c| < q$.
\end{definition}

The following result gives us a domain in which we can allocate all of the vaccines to beneficiaries. We call the sufficient conditions presented in this result by the representative name of \emph{demand law}.

\begin{theorem}[Demand law]\label{possibility result}
 Suppose that every patient $i$ is a beneficiary of at most $b$ sparse categories, and that for any sparse category $c$ it holds that $|\B_c| \geq b \cdot r_c$. Then, there exists a matching $\mu$ that assigns every vaccine to a beneficiary.
\end{theorem}
\begin{proof}
    Proof is relegated to \cref{proof of possibility result}.
\end{proof}

The proof of this result uses Hall's marriage theorem \citep{Hall35} and a double counting argument. In \cref{appendix: on demand law}, we give a pandemic-related use case of the demand law to show its practicality and give a simple general procedure to meet the sufficient conditions laid out by the demand law. Furthermore, we show that these conditions are, although not necessary, tight, whose style follows that of the maximal domain results in the literature.

The threshold reserve model does not make any assumptions about the reserves in the categories. Yet, \cref{possibility result} suggests that it might be a good idea to take the size of the categories into account when deciding how many units to allocate for each category, by giving a restricted domain in which not only both maximality axioms are satisfied but also all of the vaccines are assigned to beneficiaries. In particular, if the market designer can have impact on the number of reserves allocated in each category, then the problem of scarce resource allocation simplifies significantly. The issue of how to dereserve the excess reserves ceases to exist; we do not need to distinguish between soft and hard reserves; and the threshold reserve model does not have a use anymore. However, it is not always possible for a market designer to change the reserve numbers due to institutional rules. A recent example of this is the Indian affirmative action where the reserve numbers are pre-determined by a Supreme Court decision.\footnote{See \cite{BusinessRules23}, \cite{SY22}, or \cite{AT22} for more details.}

Thus far, we treated the desideratum of maximum in beneficiary assignments as a hard constraint. However, one other possible solution is to find an allocation that assigns the maximum number of vaccines possible, and maximizes the number of beneficiaries under that constraint. That is, we find an allocation that is maximum in beneficiary assignment among the allocations that are maximum in resource allocation. We call this desideratum by the illustrative name of \emph{max-in-max}.
}
{
The generality of the threshold reserve model comes from the fact that it does not impose any restrictions on the universal domain. Therefore, we focus on compromising from one of the optimality desiderata. A possible solution for overcoming the impossibility result is to compromise from the maximum in beneficiary assignment desideratum and to find an allocation that assigns the maximum number of vaccines possible, and maximizes the number of beneficiaries under that constraint. That is, we find an allocation that is maximum in beneficiary assignment among the allocations that are maximum in resource allocation. We call this desideratum by the illustrative name of \emph{max-in-max}.
}

\begin{definition}\label{def max-in-max}
A matching $\mu$ is \textbf{max-in-max} if $\mu$ is maximum in resource allocation and for any other matching $\mu'$ that is maximum in resource allocation
    \[\left|\cup_{c \in C}\left((\mu')^{-1}(c) \cap \B_c\right)\right| \leq \left|\cup_{c \in C}\left(\mu^{-1}(c) \cap \B_c\right) \right|.\]
\end{definition}

This desideratum places greater emphasis on optimizing the number of resources allocated than on beneficiary assignments. A natural question arises here as to why we do not do the opposite. The following result lies behind this normative decision.

\begin{proposition} \label{equivalence result}
For any matching $\mu$, there exists a matching $\mu'$ that is maximum in resource allocation such that $\{p \in \PP \mid \mu(p) \neq \varnothing \} \subseteq \{p \in \PP \mid \mu'(p) \neq \varnothing\}$.
\end{proposition}
\begin{proof}
    Proof is relegated to \cref{proof of equivalence result}.
\end{proof}

\cref{equivalence result} has some noteworthy consequences. It implies that for any matching $\mu$ that is maximum in beneficiary assignment, there is a matching $\mu'$ that is maximum in resource allocation and also assigns vaccines to the patients that are matched in $\mu$, possibly from categories of which the patients are not beneficiaries. This brings out the question of whether it is really important to match someone with a category of which she is a beneficiary rather than some other category, at the cost of matching fewer patients with vaccines, especially given that the patient does not have a preference between the two categories. On the other hand, policy makers might want to match more patients with the categories of which they are beneficiaries if they will reveal the numbers to the public in the end. In this matter, the max-in-max desideratum is opinionated, choosing to maximize resource allocation at the expense of fewer beneficiary assignments. The rest of the paper is dedicated to satisfying the max-in-max property in the universal domain without compromising from the axioms.

\section{The Max-in-Max Algorithm} \label{Max-in-Max Subsection}


In this section, we describe an algorithm to produce a max-in-max matching in the universal domain, thereby solving a new optimization problem. In subsequent sections, we will give a mechanism that works in the universal domain to output a matching that is max-in-max and also satisfy the axioms. 

A matching that is maximum in resource allocation can be found by using well-established algorithms such as the Hopcroft-Karp algorithm \citep{HK73}. The Max-in-Max algorithm takes as an input a matching that is maximum in resource allocation, and outputs a matching that is max-in-max. First, let us give some definitions that we use in the description of the algorithm.

For a given matching $\mu$, we call an ordered tuple of alternating patients and categories $(p_0, c_1, p_1, c_2, p_2, \cdots, c_m, p_m)$
a \textbf{patient-initiated even alternating chain in $\mu$} if \\
(1) either patient $p_0$ is unmatched under $\mu$, or $p_0 = p_m$ in which case it is a cycle;\footnote{Note that it could be that $p_0 = p_m$, in which case it is a cycle rather than a chain. However, for our purposes, it does not make a difference and we will not distinguish between them. Therefore, we still call it an alternating chain, even if it is in fact a cycle.} \\
(2) for every $1 \leq k \leq m$, patient $p_{k-1}$ is eligible for category $c_{k}$, that is $p_{k-1} \in \E_{c_k}$; \\
(3) for every $1 \leq k \leq m$, patient $p_k$ is assigned a vaccine from category $c_k$, that is $\mu(p_k) = c_k$.

\begin{figure}[h] 
\centering
\begin{tikzpicture}

    \Vertex[label = $c_1$, x=0, y=2]{c1}
    \Vertex[label = $c_2$, x=2, y=2]{c2}
    \Vertex[label = $c_3$, x=4, y=2]{c3}
    \Vertex[label = $c_4$, x=6, y=2]{c4}
    \Vertex[label = $c_5$, x=8, y=2]{c5}

    \Vertex[label = $p_0$, x=-2, y=0]{i0}
    \Vertex[label = $p_1$, x=0, y=0]{i1}
    \Vertex[label = $p_2$, x=2, y=0]{i2}
    \Vertex[label = $p_3$, x=4, y=0]{i3}
    \Vertex[label = $p_4$, x=6, y=0]{i4}
    \Vertex[label = $p_5$, x=8, y=0]{i5}

    \Edge[style = dashed](i0)(c1)
    \Edge[style = dashed](i1)(c2)
    \Edge[style = dashed](i2)(c3)
    \Edge[style = dashed](i3)(c4)
    \Edge[style = dashed](i4)(c5)

    \Edge(i1)(c1)
    \Edge(i2)(c2)
    \Edge(i3)(c3)
    \Edge(i4)(c4)
    \Edge(i5)(c5)

\end{tikzpicture}
\caption{A patient-initiated even alternating chain.}
\label{fig: an even alternating chain}
\end{figure}

Given an even alternating chain $(p_0, c_1, p_1, c_2, p_2, \cdots, c_m, p_m),$ we define the operation of \textbf{augmenting the chain in $\mu$} by creating a new matching $\mu'$ that differs from $\mu$ as follows: \\
(1) patient $p_m$ is unmatched under $\mu'$ if $p_0 \neq p_m$, \\
(2) for any $1 \leq k \leq m$, patient $p_{k-1}$ is assigned a vaccine from category $c_k$, that is $\mu'(p_{k-1}) = c_k$.

\begin{figure}[h]
\centering
\begin{tikzpicture}

    \Vertex[label = $c_1$, x=0, y=2]{c1}
    \Vertex[label = $c_2$, x=2, y=2]{c2}
    \Vertex[label = $c_3$, x=4, y=2]{c3}
    \Vertex[label = $c_4$, x=6, y=2]{c4}
    \Vertex[label = $c_5$, x=8, y=2]{c5}

    \Vertex[label = $p_0$, x=-2, y=0]{i0}
    \Vertex[label = $p_1$, x=0, y=0]{i1}
    \Vertex[label = $p_2$, x=2, y=0]{i2}
    \Vertex[label = $p_3$, x=4, y=0]{i3}
    \Vertex[label = $p_4$, x=6, y=0]{i4}
    \Vertex[label = $p_5$, x=8, y=0]{i5}

    \Edge(i0)(c1)
    \Edge(i1)(c2)
    \Edge(i2)(c3)
    \Edge(i3)(c4)
    \Edge(i4)(c5)

    \Edge[style = dashed](i1)(c1)
    \Edge[style = dashed](i2)(c2)
    \Edge[style = dashed](i3)(c3)
    \Edge[style = dashed](i4)(c4)
    \Edge[style = dashed](i5)(c5)

\end{tikzpicture}
\caption{The even alternating chain from \cref{fig: an even alternating chain} after the process of augmentation.}
\end{figure}

 Analogously, for a given matching $\mu$, we call an ordered tuple of alternating categories and patients $(c_0, p_1, c_1, p_2, c_2, \cdots, p_m, c_m)$
a \textbf{category-initiated even alternating chain in $\mu$} if \\
(1) either category $c_0$ has an unassigned vaccine in $\mu$, or $c_0 = c_m$ in which case it is a cycle; \\
(2) for every $1 \leq k \leq m$, patient $p_{k}$ is eligible for category $c_{k-1}$, that is $p_{k} \in \E_{c_{k-1}}$; \\
(3) for every $1 \leq k \leq m$, patient $p_k$ is assigned a vaccine from category $c_k$, that is $\mu(p_k) = c_k$.

The operation of augmentation is defined analogously to a patient-initiated even alternating chain. We call a tuple an \textbf{even alternating chain in $\mu$} if it is a patient-initiated or a category-initiated even alternating chain.\footnote{One can similarly define the notion of an \textbf{odd alternating chain} and the operation of augmentation for it. However, we will not use it in this algorithm, and therefore we omit this definition.}  

Let us make two observations. First, by the definition of alternating chains, the new matching $\mu'$ we obtain by augmenting the path in $\mu$ still complies with eligibility requirements. Second, augmenting an even alternating chain does not change the number of vaccines allocated to patients. Therefore, if a matching $\mu$ is maximum in resource allocation, the augmented matching $\mu'$ remains maximum in resource allocation. Next, we give one last definition before we state the algorithm. 

We define the \textbf{potential} of a patient-initiated even alternating chain in $\mu$ as
\[ \Phi_{\mu}(p_0, c_1, p_1, c_2, p_2, \cdots, c_m, p_m) = \: \bigm|\{ k \in [m] \mid p_{k-1} \in \B_{c_k}\}\bigm| - \bigm|\{ k \in [m] \mid p_{k} \in \B_{c_k}\}\bigm|.\]
Similarly, the potential of a category-initiated even alternating chain in $\mu$ is defined as
\[ \Phi_{\mu}(c_0, p_1, c_1, p_2, c_2, \cdots, p_m, c_m) = \: \bigm|\{ k \in [m] \mid p_k \in \B_{c_{k-1}}\}\bigm| - \bigm|\{ k \in [m] \mid p_{k} \in \B_{c_k}\}\bigm|.\]

\begin{figure}[h]
\centering
\begin{tikzpicture}

    \Vertex[label = $c_1$, x=0, y=2]{c1}
    \Vertex[label = $c_2$, x=2, y=2]{c2}
    \Vertex[label = $c_3$, x=4, y=2]{c3}
    \Vertex[label = $c_4$, x=6, y=2]{c4}
    \Vertex[label = $c_5$, x=8, y=2]{c5}

    \Vertex[label = $p_0$, x=-2, y=0]{i0}
    \Vertex[label = $p_1$, x=0, y=0]{i1}
    \Vertex[label = $p_2$, x=2, y=0]{i2}
    \Vertex[label = $p_3$, x=4, y=0]{i3}
    \Vertex[label = $p_4$, x=6, y=0]{i4}
    \Vertex[label = $p_5$, x=8, y=0]{i5}

    \Edge[style = dashed, color = black](i0)(c1)
    \Edge[style = dashed, color = black](i1)(c2)
    \Edge[style = dashed, color = red](i2)(c3)
    \Edge[style = dashed, color = black](i3)(c4)
    \Edge[style = dashed, color = red](i4)(c5)

    \Edge[color = red](i1)(c1)
    \Edge[color = black](i2)(c2)
    \Edge[color = black](i3)(c3)
    \Edge[color = black](i4)(c4)
    \Edge[color = black](i5)(c5)

\end{tikzpicture}
\caption{A patient-initiated even alternating chain with positive potential. Red edges represent beneficiary assignment.}
\end{figure}

With the necessary terminology established, we are ready to state the algorithm.

\begin{algorithm}[H] \label{Max-in-Max Algorithm Pseudocode}
\caption{Max-in-Max Algorithm}
\KwInput{A matching $\mu_0$ that is maximum in resource allocation}
\KwOutput{A max-in-max matching $\mu$}
$\mu \gets \mu_0$\;
\While{There is an even alternating chain $s$ in $\mu$ with positive potential}{
Augment the chain $s$ in $\mu$.
}
\KwRet $\mu$
\end{algorithm}

Note that it is not immediately clear how to find an even alternating chain with positive potential in a computationally efficient way. We describe an algorithm for that in \cref{appendix: finding augmenting path in poly time}. The correctness\footnote{\emph{Correctness of an algorithm} is a term commonly used in the computer science literature; although it is not as common in economics. It refers to the property that the algorithm correctly implements the intended solution to the problem it is designed to solve for any given input.} of the Max-in-Max algorithm hinges upon the following proposition.

\begin{theorem}\label{Max-in-Max correctness}
A matching $\mu$ that is maximum in resource allocation is also max-in-max if and only if there is no even alternating chain with positive potential in $\mu$.
\end{theorem}
\begin{proof}
    Proof is relegated to \cref{proof of max-in-max correctness}.
\end{proof}

Observe that if an even alternating chain has positive potential, then augmenting this chain in $\mu$ increases the number of beneficiary assignments while keeping the number of patients assigned a vaccine constant. Therefore, if there is such alternating chain present in matching $\mu$, it cannot be max-in-max. On the other hand, the proof of the converse statement is not as apparent, and we relegate it to the appendix.

The Max-in-Max algorithm leverages a key characteristic of the optimization problem at hand, that is, local optimality coincides with global optimality. It begins with an initial matching exogenous to the algorithm, augments even alternating chains with positive potential, and terminates when it cannot optimize any further. The Max-in-Max algorithm runs the augmenting path finder described in \cref{appendix: finding augmenting path in poly time} at most $q$ times, since that is the maximum number of beneficiary assignments we can possibly have. The augmenting path finder uses the Bellman-Ford algorithm, which runs in $O(|\PP|\cdot|\C|^2)$. Therefore, the runtime of the Max-in-Max algorithm is $O(q \cdot |\PP| \cdot |\C|^2)$, which simplifies to $O(q \cdot |\PP|)$ in regimes where the number of categories $|\C|$ is a bounded constant.

\section{The Iterative Max-in-Max Assignment Mechanism} \label{section: IMMAM}

The Max-in-Max algorithm provides a way to find a max-in-max matching. However, its outcome does not necessarily respect priorities. Moreover, similar to the Top Trading Cycles algorithm employed in kidney exchange mechanisms \citep{kidney04}, its outcome depends on the specifics of the implementation. We now describe the Iterative Max-in-Max Assignment Mechanism (IMMAM), which manages to respect priorities by repeatedly invoking the Max-in-Max algorithm. Moreover, the choice function that represents the patients who are assigned a vaccine in IMMAM is path independent (see \cref{def: path independence}). We leverage this property of IMMAM to prove a comparative statics result (\cref{thm: comparative statics}) and to speed up the algorithm in \cref{speeding up the algorithm}.

We begin by introducing some terminology that will be useful in describing the mechanism. First, for a given vaccine allocation market $(\PP, \CC, r, \pi)$, where the maximum number of resources allocated is $e$, and the number of beneficiary assignments in a max-in-max matching is $b$, we refer to the pair $(e, b)$ as the \textbf{max-in-max pair} of the market $(\PP, \CC, r, \pi)$.

Second, we formalize the notion of a market obtained by removing some patients while preserving the priority relations among the remaining ones. An \textbf{induced sub-market} of a vaccine allocation market $(\PP, \CC, r, \pi)$ is a market $(\PP', \CC', r', \pi')$ such that $\PP' \subseteq \PP$, $\CC' \subseteq \CC$, and for all $p, p' \in \PP'$, we have $p \: \pi \: p' \implies p \: \pi' \: p'$.
For any given $\PP' \subseteq \PP$ and $\CC' \subseteq \CC$ endowed with an arbitrary reserve profile $r'$, there exists a unique induced sub-market, denoted by $(\PP', \CC', r', \pi)$.

Before we give a formal description of the mechanism, fix a precedence list of categories, denoted by the strict preference relation $\vartriangleright$. For a given vaccine allocation market, the precedence list $\vartriangleright$ uniquely defines a lexicographic strict preference relation $\succ_{(\PP, \CC, \pi, \vartriangleright)}$ between patient-category pairs where the patient is eligible for the category through the following procedure: For any two pairs $(p, c)$ and $(p', c')$ such that $p \in \B_{c}, p' \in \B_{c'}$:
\begin{itemize}
    \item if $c \neq c'$, then $(p, c) \succ_{(\PP, \CC, \pi, \vartriangleright)} (p', c')$ if and only if $c \vartriangleright c'$,
    \item if $c = c'$, then $(p, c) \succ_{(\PP, \CC, \pi, \vartriangleright)} (p', c')$ if and only if $p \: \pi_{c} \: p'$.
\end{itemize}
The precedence list $\vartriangleright$ will be an input to the mechanism together with the vaccine allocation market itself. 

\begin{algorithm}[H]
\caption{Iterative Max-in-Max Assignment Mechanism (IMMAM)} \label{iterative mechanism}
\KwInput{a vaccine allocation market $(\PP_0,\C_0, r_0,\pi_0)$, and a precedence list $\vartriangleright$ of categories}
\KwOutput{a matching $\mu$}
Using the Max-in-max algorithm, compute the max-in-max pair $(e_0, b_0)$ of the vaccine allocation market $(\PP_0,\C_0, r_0,\pi_0)$;\\
Start with the empty matching $\mu \leftarrow \mu_0 := \varnothing$, the vaccine allocation market $(\PP,\C, r,\pi) \leftarrow (\PP_0,\C_0, r_0,\pi_0)$, and the max-in-max pair $(e_0, b_0)$; \\
\For{$(p, c)$ in $\{(p, c) \mid p \in \E_{c}\}$ with the associated order of iteration $\succ_{(\PP_0, \CC_0, \pi_0, \vartriangleright)}$}{
\If{\emph{$p \not \in \PP$ \textbf{or} $r_c = 0$}}
{Skip to the next pair.}
Tentatively decrease $r_c$ by one, and remove patient $p$ from the market to obtain an induced vaccine allocation market $(\PP \setminus \{p\},\C, (r_{-c}, r_c - 1),\pi)$; \\
Using the Max-in-Max algorithm, compute the max-in-max pair $(e', b')$ of the induced market; \\
\If{\emph{$e' = e-1$ \textbf{and} ($b' = b$ \textbf{or} ($p \in \B_c$ \textbf{and} $b' = b-1$))}}
{
    Assign $\mu \leftarrow \mu \cup \{(p, c)\}$; \\
    Assign $e \leftarrow e'$ and $b \leftarrow b'$; \\
    Assign $(\PP,\C, r,\pi) \leftarrow (\PP \setminus \{p\},\C, (r_{-c}, r_c - 1),\pi)$;
}
}
\KwRet $\mu$;
\end{algorithm}

As IMMAM iterates over patient-category pairs, the set of available matchings gradually shrinks. This is because, for each pair, the mechanism makes a permanent decision about whether the patient will be assigned a vaccine from the corresponding category—while preserving all remaining possibilities for unprocessed pairs. By the time the mechanism terminates, a decision has been made for every pair, and only one matching remains in the set. Hence, we call the \emph{if} condition ($e' = e-1 \text{ and } b' = b \text{ or (}(p \in \B_c \text{ and } b' = b-1\text{)}$) in the description of IMMAM the \textbf{permanent assignment condition}. This condition ensures, by construction, that a pair $(p,c)$ is matched if and only if there remains a max-in-max matching in the set of available matchings, thereby guaranteeing that the outcome will be max-in-max.


 \begin{theorem}\label{thm: IMMAM properties}
     IMMAM complies with eligibility requirements, respects priorities, admits universal domain, and is max-in-max.
 \end{theorem}
 \begin{proof}
     Proof is relegated to \cref{proof of IMMAM properties}.
 \end{proof}

 Note that the max-in-max property of IMMAM implies that it is maximum in resource allocation, by definition. 
 
 Keeping the profile of categories, reserves, and priorities fixed and varying the set of patients IMMAM defines a choice function $C^{\vartriangleright}_{(\PP, \C, r,\pi)} : 2^{\PP} \to 2^{\PP}$ on the set of patients $\PP$ such that $C^{\vartriangleright}_{(\PP, \C, r,\pi)}(X) \subseteq X$ for any $X \subseteq \PP$. Here, $\PP$ represents the of all individuals who are potential market participants, whereas $X$ is the set of patients who actually participate in the market.\footnote{Note that for any $\PP' \subseteq \PP$, the choice function $C^{\vartriangleright}_{(\PP', \C, r,\pi)}(\cdot)$ defined for the induced sub-market $(\PP', \CC, r,\pi)$ coincides with $C^{\vartriangleright}_{(\PP, \C, r,\pi)}(\cdot)$ on any set of patients in the power set $2^{\PP'}$ (which we formally make note of in \cref{equivalence of choices in induced markets}). Thus, this choice function is not inherently specific to the set of patients $\PP$. However, in order for the choice function to be well defined, we must specify the priority profile, which requires defining the set of patients $\PP$ over which those priorities are expressed. As such, the reader may find it helpful to interpret $\PP$ here as the universal set of patients, including all individuals, regardless of whether they participate in the market. In this view, we may think of the choice function as a single object, where a possible input $X$ to the choice function represents the set of market participants, and ignore the index in the notation.} 
The set of chosen patients changes as the set of market participants varies, exhibiting desirable comparative statics properties.

\begin{theorem}\label{thm: comparative statics}
\begin{enumerate}
    \item  If a patient leaves the market, everyone else is weakly better off. That is, for any $p \in X \subseteq \PP$, we have $C^{\vartriangleright}_{(\PP, \C, r,\pi)}(X) \subseteq C^{\vartriangleright}_{(\PP, \C, r,\pi)}(X \setminus \{p\}) \cup \{p\}$.
    
    \item If there is an additional vaccine in any one of the categories, then every patient is weakly better off. That is, for any $X \subseteq \PP$ and $c \in \C$, we have $C^{\vartriangleright}_{(\PP, \C, r,\pi)}(X) \subseteq C^{\vartriangleright}_{(\PP, \C, (r_{-c}, r_c+1),\pi)}(X)$.
\end{enumerate}
\end{theorem}
\begin{proof}
    Proof is relegated to \cref{proof: comparative statics}.
\end{proof}
The proof of this result uses the fact that the choice function $C^{\vartriangleright}_{(\PP, \C, r,\pi)}(\cdot)$ satisfies \emph{path independence.}

\begin{definition}\label{def: path independence}
    A choice function $C(\cdot)$ on the set of patients satisfies \textbf{path independence} if for any $X, Y \subseteq \PP$, we have $C(X \cup Y) = C(C(X) \cup C(Y))$.
\end{definition}

Path independence says that if we need to include new patients in the mechanism after already running it, keeping the priorities between any pair of existing patients within each category the same, it suffices to simply combine them with the set of chosen patients and re-run the algorithm.

\begin{theorem}\label{thm: path independence}
    The choice function $C^{\vartriangleright}_{(\PP, \C, r,\pi)} : 2^{\PP} \to 2^{\PP}$ defined by IMMAM is path independent.
\end{theorem}
\begin{proof}
Proof is relegated to \cref{section: proof of path independence}.
\end{proof}

The first statement in \cref{thm: comparative statics} follows directly from path independence,\footnote{The first statement in \cref{thm: comparative statics} is equivalent to the \emph{substitutability} condition, which is weaker than path independence (\cref{path independence = consistency and substitutability condition}). In fact, when proving path independence, we first prove substitutability separately (\cref{IMMAM is substitutable}).} and the second statement builds on the first in its proof. Path independence is important not only for its comparative statics implications but also for two additional reasons in the context of this paper.

First, although we do not explore it further within the scope of this paper, path independence guarantees the existence of a stable outcome in settings with multiple institutions allocating vaccines using the same choice function, where patients have preferences over institutions \citep{CY17}. These institutions can represent hospitals or pharmacies, as well as different vaccine brands, which virtually allows patients to have preferences over vaccines by separating their distributions.

Second, path independence has algorithmic implications that we leverage for efficiency. For any set of patients, the algorithm selects at most $q$ of them, where $q$ is the total number of vaccines available. When vaccines are a scarce resource, this bounded selection implies that we can reduce computational effort in subsequent iterations if the set of patients changes. In the second part of the following section, we exploit this observation to improve the runtime of IMMAM, highlighting the practical algorithmic benefits of path independence.

\subsection{Speeding up IMMAM} \label{speeding up the algorithm}
 The main expense in IMMAM is running the Max-in-Max algorithm repeatedly. In particular, it is run at most $|\PP| \cdot |\C|$ times. Thus, its worst-case runtime is $O(q \cdot |\PP|^2\cdot|\C|^3)$. In this section, we reduce this runtime to $O(q \cdot |\PP| \cdot |\CC|^4)$ by applying two steps of optimizations. Since we are in a regime where $|\PP| \gg q \gg |\CC|$ due to scarcity, and $|\CC|$ is a small constant in practice, this optimization arguably reduces the (parameterized) runtime from quadratic to linear in the number of patients. In our setup, it is natural to assume that the number of patients is much greater than the number of vaccines available since vaccines are treated as a scarce resource. Moreover, the number of vaccines is greater than the number of categories, since each category must have a vaccine. In all practical applications of reserve systems, the number of categories is only a handful, and does not necessarily increase by the number of unit resources available, thus may be considered $O(1)$. To preserve generality, we still consider it as a variable when analyzing runtime.

\subsubsection{First Optimization: IMMAM with Memory (IMMAM-M)}

Our first optimization involves keeping track of the initial max-in-max matching and augmenting it in place instead of finding a new max-in-max matching at each iteration. In IMMAM, we repeatedly compute the max-in-max matching from scratch. However, it is possible to reuse the max-in-max matching from the previous iteration in the induced sub-market in each iteration. We do so by finding the augmenting paths starting from the last matching. This optimization enables us to reduce the maximum number of augmentations in the Max-in-Max algorithm from $q$ to $|\C|$, where $q$ is the total number of vaccines in the market. Therefore, the Max-in-Max algorithm will need to run the augmenting path finder at most $|\C|$ times, which reduces the runtime of each iteration from $O(q \cdot |\PP|\cdot|\C|^2)$ to $O(|\PP|\cdot|\C|^3)$. Thus, the total runtime becomes $O(|\PP|^2\cdot|\C|^4)$. We call this optimized version \emph{IMMAM with Memory (IMMAM-M)}. A detailed pseudo-code for IMMAM-M can be found in \cref{appendix: IMMAM-M pseudo}.

\subsubsection{Second Optimization: IMMAM-M with Bootstrapping (IMMAM-MB)}

Finally, the further optimized version bootstraps IMMAM-M using the fact that the choice function corresponding to IMMAM is path independent (\cref{thm: path independence}) and that the algorithm chooses at most $q$ patients regardless of the input size. We call this version \emph{IMMAM-M with Bootstrapping (IMMAM-MB)}, and it runs in $O(q\cdot|\PP|\cdot|\C|^4)$ time. This optimization technique signifies the importance of path independence in algorithmic complexity considerations, and can be generalized to any path independent choice rule whose output size is bounded above by a constant.

\begin{algorithm}[H]
\caption{IMMAM-M with Bootstrapping (IMMAM-MB)} \label{recursive mechanism}
\KwInput{a vaccine allocation market $(\PP,\C, r,\pi)$, and a precedence list $\vartriangleright$ of categories}
\KwOutput{a matching $\mu$}
\If{$|\PP| \leq 2q$}{
    Run IMMAM-M on $((\PP,\C, r,\pi), \vartriangleright)$, and \KwRet the resulting matching;
}
\Else{
    Split the patients into two arbitrary sets $\PP_1$ and $\PP_2$ of roughly the same size; \\
    Run IMMAM-MB on $((\PP_1,\C, r,\pi), \vartriangleright)$ and let the set of patients who are assigned a vaccine be $\PP_1^{\text{(chosen)}}$; \\
    Run IMMAM-MB on $((\PP_2,\C, r,\pi), \vartriangleright)$ and let the set of patients who are assigned a vaccine be $\PP_2^{\text{(chosen)}}$; \\
    Run IMMAM-M on  $((\PP_1^{\text{(chosen)}} \cup \PP_2^{\text{(chosen)}}, \C, r,\pi), \vartriangleright)$, and \KwRet the resulting matching;
}
\KwRet $\mu$;
\end{algorithm}

Inherently, IMMAM-MB splits the patients into groups of $2q$, and runs IMMAM-M on each, where the resulting groups will each have size at most $q$. Then, it combines these groups pairwise, and runs the algorithm on the new groups of size $2q$, and continues recursively. Throughout the procedure, we run IMMAM-M fewer than $|\PP|/q$ times, each of which is on a patient set of size at most $2q$. Therefore, the runtime is $O(q\cdot|\PP|\cdot|\C|^4)$, which corresponds to a factor of $|\PP|/q$ improvement on top of IMMAM-M, bringing the overall improvement over IMMAM to be a factor of $|\PP|/|\CC|$.

Finally, the following result is needed to conclude that our optimizations did not change the resulting matching.
\begin{proposition} \label{thm: outcome equivalence}
   IMMAM-MB and IMMAM are outcome-equivalent. 
\end{proposition}
\begin{proof}
    Proof is relegated to \cref{proof: outcome equivalence}.
\end{proof}
Two mechanisms being outcome-equivalent means that the resulting matching is the same for any given input. Therefore, any property IMMAM satisfies in terms of its outcome, so does IMMAM-MB. The proof of this result uses the fact that IMMAM is path independent, and it is relegated to the appendix.

\section{Related Literature on Theory and Vaccine Allocation} \label{section: review}
\subsection{Related Literature} \label{subsection: related literature}
The threshold reserve model we use differs from the existing smart reserves frameworks in two main ways. First, previous literature assumes either all \emph{soft} or all \emph{hard} reserves. A soft reserve corresponds to having every patient above the beneficiary threshold in our model in every given category, while a hard reserve requires eligibility threshold to be the same as the beneficiary threshold, i.e., no non-beneficiary is eligible in any given category. The model developed in \cite{AT22} allows soft reserves in a single category while keeping the reserves in all the other categories as hard reserves. However, their model cannot be trivially extended to allow multiple categories with soft reserves, and ad-hoc to the problem of Indian affirmative action. It also does not accommodate the presence of overlapping categories. Our model allows some categories to follow soft reserves and some categories to follow hard reserves. Furthermore, it also allows any given category to be neither soft nor hard reserve, but a mixture of both.

Second, the previously used models mostly assume the presence of a baseline priority ordering. This enforces the priority orderings of different categories to have significant inter-dependencies, which might be detrimental especially taking into account the fact that the main use of reserve systems is to accommodate different groups at the same time without having to compare two incomparable segments.

The one paper that analyzes a model that does not assume baseline priority ordering is \cite*{PSUY20}. However, their model corresponds to a hard reserve. In the same paper, they propose another model with baseline priority orderings, for which they describe a smart reserve matching mechanism. The mechanism they describe is an extension of the Meritorious Horizontal Choice Rule (MHCR) introduced in \cite{SY22}.\footnote{This paper subsumed two working papers of the authors. Meritorious Horizontal Choice Rule was first introduced in \cite{SY20} as Horizontal Envelope Choice Rule.} IMMAM resembles MHCR in its iterative structure; IMMAM iterates over patient-category pairs, whereas MHCR iterates over patients based on a baseline priority ordering.

Reserve systems are part of a larger literature regarding affirmative action processes. \cite{AS03} studied the policies imposing hard quotas on the number of admitted majority students in the context of school choice. Later, \cite{K12} showed that imposing hard quotas on majority students can hurt minority students in some cases. To overcome these possibly detrimental effects of the quota systems, \cite*{Hafalir...13} introduced reserve systems. The difference between hard and soft reserves were discussed in \cite*{EHYY14}, showing that using soft reserves when applicable may lead to more efficient outcomes. \cite{EY15} introduced \emph{minimum guarantee choice rule} in the absence of overlapping categories with its axiomatic characterization.

\cite{KS16} introduced the \emph{matching with slot-specific priorities} model, which was later on adapted in several papers studying real-life applications of reserve systems with non-overlapping categories such as \cite*{DKPS18} for school choice in Boston, and \cite*{DPS20} in the context of school choice in Chicago. Some recent papers that study non-overlapping reserves, all in the context of Indian affirmative action, are by \cite{AT17, AT20, AT22, AT23}.

The study of overlapping reserve categories is a more recent development in the field. Some papers include \cite*{gonczarowski2020matching} in the context of diversity requirements of pre-military gap year programs in Israel, \cite{AB21} in the context of affirmative action policies for Brazilian colleges, \cite{SY22} for Indian affirmative action, and \cite{PSUY20} for medical rationing and vaccine allocation. A theory of generic reserve systems with overlapping categories was developed by \cite{PSUY20}. Unlike our model, the model in their paper does not incorporate the smartness constraints and the option of having two thresholds. Rather, they only have a set of agents who are eligible for each category.

Our Max-in-Max algorithm contributes to the fields of matching theory and combinatorial optimization. In the Max-in-Max algorithm, we use the augmenting paths technique. Some well-known algorithms using the augmenting paths technique include the Hopcroft-Karp algorithm \citeyearpar{HK73}, the Edmond's Blossom algorithm \citeyearpar{E65}, the Ford-Fulkerson algorithm \citeyearpar{FF56} for finding a maximum matching, and the Hungarian algorithm \citep{Kuhn55, Munkres57} for finding a maximum one-to-one weighted matching. Note that we use the Hopcroft-Karp algorithm to find a matching that is maximum in resource allocation as input to Max-in-Max. A more recent strand of the literature builds on the Hungarian algorithm to compute one-to-many \citep{Many-to-One-KM} and many-to-many \citep{Many-to-many-KM} weighted matchings in a computationally efficient manner.

Note that algorithms mentioned in the above paragraph are used to maximize over a single objective. Our Max-in-Max algorithm also contributes to the field of multi-objective optimization problems. The max-in-max desideratum is a \emph{lexicographic} optimization objective, and the Max-in-Max algorithm brings a tailored solution to this lexicographic maximum matching problem.

\subsection{Vaccine Allocation}\label{application: vaccine}

Contextualizing our results, we adapt the language of vaccine allocation problem, closely aligned with the language developed by \cite*{PSUY20}, to be consistent with the recent reserve systems literature. We also think that following this language rather than using a more abstract one makes it easier to convey the general idea of our paper. In this section, we provide background regarding the problem of vaccine allocation and argue that our proposed framework has the potential to work better.

With the onset of the COVID-19 pandemic, priority systems have been implemented to allocate vaccines to individuals. In a priority system, patients are linearly ordered and the vaccines are given to people following this order. Such mechanisms have two widely recognized issues.

First, it effectively requires the resolution of various difficult ethical dilemmas. For example, in a priority system, one must decide whether to prioritize infants or the elderly. Life cycle principle suggests that we should prioritize the infants out of fairness since they have not gone through the cycles of life yet. On the other hand, we may want to prioritize the elderly in order to save the most lives, since their immune system is more vulnerable. Putting infants and elderly in a single line requires some trade off between these two conflicting ethical principles.

One possible solution to the problem of conflicting ethical principles is to use a mixture of them, known as the \emph{multi-principle} approach. A common way to achieve this is through the use of \emph{priority point systems}, in which patients are scored based on different criteria and their order is decided based on their total scores. Such solutions end up comparing incomparable groups. For example, the \emph{multi-principle point system} developed by \cite*{White...2009} ends up comparing different age groups by the virtue of using Sequential Organ Failure Assessment (SOFA) scores. However, according to Massachusetts guidelines, the indicators that are used to measure the health of adults are not reliable when applied to infants (\citeyear{MassGuideline20}).\footnote{For more details regarding SOFA scores, see \cite{SOFA96}. For more details and discussion on the differences between approaches for adult and pediatric patients, see pages 17-20 in the Massachusetts Planning Guidance for the COVID-19 Pandemic (\citeyear{MassGuideline20}). For more discussion on the issues regarding widely used medical resource allocation systems, see \cite{PSUY20} and \cite*{ABDK23}.} 

To combat these issues, \cite{PSUY20} suggested the use of reserve systems. In the model they propose, there are different categories with their own priority orderings, and a set of beneficiaries for each category. They assume that the vaccines are identical and safe for everybody. In this reserve system, each category accommodates a different ethical value, which enables policymakers to avoid making decisions on ethical dilemmas and compare incomparable patients. In particular, the authors proposed two different solutions: Patient-proposing Deferred Acceptance and Smart Reserve Matching Algorithm. The main trade-off between their proposed solutions are as follows. The former does not impose any restrictions on the priority orderings of different categories, but fails to efficiently allocate the maximum number of vaccines possible to beneficiaries of the groups. Conversely, the latter is maximum in beneficiary assignment but requires significant interdependencies (existence of a baseline priority ordering) between the priority orderings of different groups.  

The solution we propose contributes to the existing solution to the vaccine allocation problem in two different ways. First, our mechanism combines the best aspects of both mechanisms described by \cite{PSUY20}. It does not assume any interdependencies between categories, and optimizes over the number of vaccines assigned to higher tiers. Second, our model allows three different tiers, which might be necessary in scenarios where we do not want to waste any excess vaccine from a category but also not every vaccine is safe for every patient, as we have already discussed further in \cref{intro} and \cref{model}.

\section{Conclusion} \label{conclusion}

In this paper, we introduce a comprehensive threshold reserve model for smart reserve matchings with overlapping categories that have arbitrary priority orderings exogenous to the model. We presented characterization results in the landscape of our model. Our first result shows the impossibility of satisfying both optimality constraints at the same time. \ifthenelse{\boolean{demand_law_on}}{  We introduce a demand law, showing that if the demand in each category is high enough, then we can allocate all of the units to beneficiaries. To the best of our knowledge, this is the first theoretical result that offers guidance on how to distribute the resources to different reserve categories. An interesting future direction here is to incorporate data to observe how often the demand law conditions are met in the settings where the reserve numbers are pre-determined.}{} Building on the impossibility result, we introduced a new objective called max-in-max. This approach requires the designer to first maximize the total number of resources allocated, and then, within that constraint, optimize the number of beneficiary assignments.

Next, we first presented the Max-in-Max algorithm for finding a max-in-max matching, which solves the lexicographic optimization problem at hand. Then, we introduced the Iterative Max-in-Max Assignment Mechanism (IMMAM) to obtain a max-in-max matching that respects priorities. This mechanism applies the Max-in-Max algorithm iteratively by processing category–patient pairs in a well-specified order.

Varying the set of patients in the vaccine allocation market, the chosen set of patients defines a choice function, which we showed to be path independent. Path independence implies that in presence of multiple institutions distributing vaccines over which the patients may have preferences, there always exists a stable matching. We also used IMMAM's path independence in proving useful comparative statics results, answering the question of what happens if a patient leaves the market or if the vaccine production increases. Finally, we used path independence for improving the runtime of our mechanism to $O(q\cdot|\PP|\cdot|\C|^4)$, which reduces to $O(q\cdot|\PP|)$ in the cases where the number of categories is assumed to be a small constant.

Our suggested solution provides policymakers with flexibility in affirmative action processes and resource allocation, equipping them with a more general model without imposing hard restrictions on priorities and thresholds. The threshold reserve model the paper develops encompasses several previously suggested solutions to allocation problems in different narrower domains. In addition, it enables them to not compare incomparable segments of the population and not sacrifice certain ethical values in the system they design, not requiring them to put everybody in one line as they would have to do in a baseline priority setting. Our mechanism results in an optimal solution that maximizes the number of units utilized, and serves the primary purpose of categories by further optimizing over the number of beneficiaries receiving a unit.

\bibliography{ref.bib}

@article{PSUY20,
  
  author = {Pathak, Parag A. and S\"{o}nmez, Tayfun and \"{U}nver, M. Utku and Yenmez, M. Bumin},
  
  title = {Fair Allocation of Vaccines, Ventilators and Antiviral Treatments: Leaving No Ethical Value Behind in Health Care Rationing},
  
  journal = {Management Science, forthcoming},
  
  year = {2023},
  
}

@article{Hall35,
author = {Hall, Philip},
title = {On Representatives of Subsets},
journal = {Journal of the London Mathematical Society},
volume = {s1-10},
number = {1},
pages = {26-30},
doi = {https://doi.org/10.1112/jlms/s1-10.37.26},
url = {https://londmathsoc.onlinelibrary.wiley.com/doi/abs/10.1112/jlms/s1-10.37.26},
eprint = {https://londmathsoc.onlinelibrary.wiley.com/doi/pdf/10.1112/jlms/s1-10.37.26},
year = {1935}
}

@article{Dulmage-Mendelsohn58, title={Some Generalizations of the Problem of Distinct Representatives}, volume={10}, DOI={10.4153/CJM-1958-027-8}, journal={Canadian Journal of Mathematics}, publisher={Cambridge University Press}, author={Mendelsohn, N. S. and Dulmage, A. L.}, year={1958}, pages={230–241}}

@article{White...2009,
title = {Who should receive life support during a public health emergency? Using ethical principles to improve allocation decisions},
DOI= {https://doi.org/10.7326/0003-4819-150-2-200901200-00011},
journal = {Annals of Internal Medicine},
author = {White, D. B. and Katz, M. H. and Luce, J. M. and Lo, B.},
year = {2009},
volume = {150(2)},
pages = {132–138} }

@article{SOFA96,
author = {Vincent, J. L. and Moreno, R. and Takala, J. and Willatts, S. and De Mendonça, A. and Bruining, H. and Reinhart, C. K. and Suter, P. M. and Thijs, L. G.}, 
year = {1996},
title = {The SOFA (Sepsis-related Organ Failure Assessment) score to describe organ dysfunction/failure. On behalf of the Working Group on Sepsis-Related Problems of the European Society of Intensive Care Medicine},
journal ={Intensive Care Medicine},
volume = {22(7)}, 
pages = {707–710},
DOI = {https://doi.org/10.1007/BF01709751}}

@article{MassGuideline20,
author = {{The Commonwealth of Massachusetts Executive Office of Health and Human Services Department of Public Health}}, 
year = {2020},
title = {Crisis Standards of Care Planning Guidance for the COVID-19 Pandemic},}

@article{ABDK23,
    author = {Akbarpour, Mohammad and Budish, Eric and Dworczak, Piotr and Kominers, Scott Duke},
    title = {An Economic Framework for Vaccine Prioritization*},
    journal = {The Quarterly Journal of Economics},
    volume = {139},
    number = {1},
    pages = {359-417},
    year = {2023},
    month = {06},
    issn = {0033-5533},
    doi = {10.1093/qje/qjad022},
    url = {https://doi.org/10.1093/qje/qjad022},
    eprint = {https://academic.oup.com/qje/article-pdf/139/1/359/55124604/qjad022.pdf},
}

@article{McCabe...89,
title = {Designing ‘smart’ computer-assisted markets: An experimental auction for gas networks},
journal = {European Journal of Political Economy},
volume = {5},
number = {2},
pages = {259-283},
year = {1989},
issn = {0176-2680},
doi = {https://doi.org/10.1016/0176-2680(89)90049-9},
url = {https://www.sciencedirect.com/science/article/pii/0176268089900499},
author = {Kevin A. McCabe and Stephen J. Rassenti and Vernon L. Smith}
}

@article{McCabe...91,
author = {Kevin A. McCabe  and Stephen J. Rassenti  and Vernon L. Smith },
title = {Smart Computer-Assisted Markets},
journal = {Science},
volume = {254},
number = {5031},
pages = {534-538},
year = {1991},
doi = {10.1126/science.254.5031.534},
URL = {https://www.science.org/doi/abs/10.1126/science.254.5031.534},
eprint = {https://www.science.org/doi/pdf/10.1126/science.254.5031.534}}

@article{Roth07,
author = {Alvin E. Roth},
title = {The Art of Designing Markets},
year = {2007},
journal = {Harvard Business Review}
}

@article{kidney07,
Author = {Roth, Alvin E. and S\"{o}nmez, Tayfun and \"{U}nver, M. Utku},
Title = {Efficient Kidney Exchange: Coincidence of Wants in Markets with Compatibility-Based Preferences},
Journal = {American Economic Review},
Volume = {97},
Number = {3},
Year = {2007},
Month = {June},
Pages = {828-851},
DOI = {10.1257/aer.97.3.828},
URL = {https://www.aeaweb.org/articles?id=10.1257/aer.97.3.828}}

@article{kidney04,
author = {Roth, Alvin E. and S\"{o}nmez, Tayfun and \"{U}nver, M. Utku},
title = {Kidney Exchange},
journal = {The Quarterly Journal of Economics},
year = {2004},
volume = {119},
pages = {457- 488},
DOI = {https://doi.org/10.1162/0033553041382157}}

@article{SY20,
author = {Tayfun S\"{o}nmez and M. Bumin Yenmez},
year = {2020},
title = {Affirmative Action with Overlapping Reserves},
journal = {Boston College Working Paper 990}}

@article{Hafalir...13,
author = {Hafalir, Isa E. and Yenmez, M. Bumin and Yildirim, Muhammed A.},
title = {Effective affirmative action in school choice},
journal = {Theoretical Economics},
volume = {8},
number = {2},
pages = {325-363},
doi = {https://doi.org/10.3982/TE1135},
url = {https://onlinelibrary.wiley.com/doi/abs/10.3982/TE1135},
eprint = {https://onlinelibrary.wiley.com/doi/pdf/10.3982/TE1135},
year = {2013}
}

@article{SY22,
author = {S\"{o}nmez, Tayfun and Yenmez, M. Bumin},
title = {Affirmative Action in India via Vertical, Horizontal, and Overlapping Reservations},
journal = {Econometrica},
volume = {90},
number = {3},
pages = {1143-1176},
doi = {https://doi.org/10.3982/ECTA17788},
url = {https://onlinelibrary.wiley.com/doi/abs/10.3982/ECTA17788},
eprint = {https://onlinelibrary.wiley.com/doi/pdf/10.3982/ECTA17788},
year = {2022}
}

@article{Glazer99,
  title={Winning in smart markets},
  author={Rashi Glazer},
  journal={Journal of Interactive Marketing},
  year={1999},
  volume={13},
  pages={2 - 4}
}

@article{BGK10,
 ISSN = {10477047, 15265536},
 URL = {http://www.jstor.org/stable/23015637},
 author = {Martin Bichler and Alok Gupta and Wolfgang Ketter},
 journal = {Information Systems Research},
 number = {4},
 pages = {688--699},
 publisher = {INFORMS},
 title = {Research Commentary: Designing Smart Markets},
 urldate = {2023-04-18},
 volume = {21},
 year = {2010}
}

@article{AB21,
Author = {Ayg\"{u}n, Orhan and Bó, Inácio},
Title = {College Admission with Multidimensional Privileges: The Brazilian Affirmative Action Case},
Journal = {American Economic Journal: Microeconomics},
Volume = {13},
Number = {3},
Year = {2021},
Month = {August},
Pages = {1-28},
DOI = {10.1257/mic.20170364},
URL = {https://www.aeaweb.org/articles?id=10.1257/mic.20170364}}

@article{K12,
title = {School choice: Impossibilities for affirmative action},
journal = {Games and Economic Behavior},
volume = {75},
number = {2},
pages = {685-693},
year = {2012},
issn = {0899-8256},
doi = {https://doi.org/10.1016/j.geb.2012.03.003},
url = {https://www.sciencedirect.com/science/article/pii/S0899825612000437},
author = {Fuhito Kojima}
}

@article{AS03,
Author = {Abdulkadiroğlu, Atila and S\"{o}nmez, Tayfun},
Title = {School Choice: A Mechanism Design Approach },
Journal = {American Economic Review},
Volume = {93},
Number = {3},
Year = {2003},
Month = {June},
Pages = {729-747},
DOI = {10.1257/000282803322157061},
URL = {https://www.aeaweb.org/articles?id=10.1257/000282803322157061}}

@article{EY15,
Author = {Echenique, Federico and Yenmez, M. Bumin},
Title = {How to Control Controlled School Choice},
Journal = {American Economic Review},
Volume = {105},
Number = {8},
Year = {2015},
Month = {August},
Pages = {2679-94},
DOI = {10.1257/aer.20130929},
URL = {https://www.aeaweb.org/articles?id=10.1257/aer.20130929}}

@article{KS16,
author = {Kominers, Scott Duke and S\"{o}nmez, Tayfun},
title = {Matching with slot-specific priorities: Theory},
journal = {Theoretical Economics},
volume = {11},
number = {2},
pages = {683-710},
doi = {https://doi.org/10.3982/TE1839},
url = {https://onlinelibrary.wiley.com/doi/abs/10.3982/TE1839},
eprint = {https://onlinelibrary.wiley.com/doi/pdf/10.3982/TE1839},
year = {2016}
}

@article{DKPS18,
author = {Dur, Umut and Kominers, Scott Duke and Pathak, Parag A. and S\"{o}nmez, Tayfun},
title = {Reserve Design: Unintended Consequences and the Demise of Boston’s Walk Zones},
journal = {Journal of Political Economy},
volume = {126},
number = {6},
pages = {2457-2479},
year = {2018},
doi = {10.1086/699974},

URL = { 
    
        https://doi.org/10.1086/699974
    
    

},
eprint = { 
    
        https://doi.org/10.1086/699974
    
    

}
}

@Article{DPS20,
  author={Dur, Umut and Pathak, Parag A. and S\"{o}nmez, Tayfun},
  title={{Explicit vs. statistical targeting in affirmative action: Theory and evidence from Chicago's exam schools}},
  journal={Journal of Economic Theory},
  year=2020,
  volume={187},
  number={C},
  pages={},
  month={},
  doi={10.1016/j.jet.2020.104996},
  url={https://ideas.repec.org/a/eee/jetheo/v187y2020ics0022053118302801.html}
}

@article{AT17,
Author = {Ayg\"{u}n, Orhan and Turhan, Bertan},
Title = {Large-Scale Affirmative Action in School Choice: Admissions to IITs in India},
Journal = {American Economic Review},
Volume = {107},
Number = {5},
Year = {2017},
Month = {May},
Pages = {210-13},
DOI = {10.1257/aer.p20171049},
URL = {https://www.aeaweb.org/articles?id=10.1257/aer.p20171049}}

@article{AT20,
title = {Dynamic reserves in matching markets},
journal = {Journal of Economic Theory},
volume = {188},
pages = {105069},
year = {2020},
issn = {0022-0531},
doi = {https://doi.org/10.1016/j.jet.2020.105069},
url = {https://www.sciencedirect.com/science/article/pii/S0022053120300661},
author = {Orhan Ayg\"{u}n and Bertan Turhan}
}

@article{EHYY14,
title = {School choice with controlled choice constraints: Hard bounds versus soft bounds},
journal = {Journal of Economic Theory},
volume = {153},
pages = {648-683},
year = {2014},
issn = {0022-0531},
doi = {https://doi.org/10.1016/j.jet.2014.03.004},
url = {https://www.sciencedirect.com/science/article/pii/S0022053114000301},
author = {Lars Ehlers and Isa E. Hafalir and M. Bumin Yenmez and Muhammed A. Yildirim}
}

@article{AT22,
author = {Ayg\"{u}n, Orhan and Turhan, Bertan},
title = {How to De-Reserve Reserves: Admissions to Technical Colleges in India},
journal = {Management Science},
volume = {0},
number = {0},
year = {2022},
doi = {10.1287/mnsc.2022.4566},

URL = { 
    
        https://doi.org/10.1287/mnsc.2022.4566
    
    

},
eprint = { 
    
        https://doi.org/10.1287/mnsc.2022.4566
    
    

}
}

@article{AT23,
author = {Ayg\"{u}n, Orhan and Turhan, Bertan},
title = {Priority design for engineering colleges in India},
journal = {Indian Economic Review},
year = {2023},
doi = {10.1007/s41775-023-00162-3},

URL = { 
    
        https://doi.org/10.1007/s41775-023-00162-3  

},
}

@article{HK73,
author = {Hopcroft, John E. and Karp, Richard M.},
title = {An $n^{5/2}$ Algorithm for Maximum Matchings in Bipartite Graphs},
journal = {SIAM Journal on Computing},
volume = {2},
number = {4},
pages = {225-231},
year = {1973},
doi = {10.1137/0202019},

URL = { 
    
        https://doi.org/10.1137/0202019
    
    

},
eprint = { 
    
        https://doi.org/10.1137/0202019
    
    

}
}

@article{FF56,
  title={Maximal Flow Through a Network},
  author={Lester Randolph Ford and Delbert Ray Fulkerson},
  journal={Canadian Journal of Mathematics},
  year={1956},
  volume={8},
  pages={399 - 404}
}

@article{E65, 
title={Paths, Trees, and Flowers}, 
volume={17}, 
DOI={10.4153/CJM-1965-045-4}, 
journal={Canadian Journal of Mathematics},
publisher={Cambridge University Press}, 
author={Edmonds, Jack}, 
year={1965}, 
pages={449–467}}

@website{Pfizer-for-adolescents,
author = {FDA},
title = {Coronavirus (COVID-19) Update: FDA Authorizes Pfizer-BioNTech COVID-19 Vaccine for Emergency Use in Adolescents in Another Important Action in Fight Against Pandemic},
year = {2021},
URL = {https://www.fda.gov/news-events/press-announcements/coronavirus-covid-19-update-fda-authorizes-pfizer-biontech-covid-19-vaccine-emergency-use},
}

@article{gonczarowski2020matching,
      title={Matching for the Israeli "Mechinot" Gap-Year Programs: Handling Rich Diversity Requirements}, 
      author={Yannai A. Gonczarowski and Lior Kovalio and Noam Nisan and Assaf Romm},
      year={2020},
      journal = {arXiv preprint}
}

@article{Bellman58,
 ISSN = {0033569X, 15524485},
 URL = {http://www.jstor.org/stable/43634538},
 author = {Richard Bellman},
 journal = {Quarterly of Applied Mathematics},
 number = {1},
 pages = {87--90},
 publisher = {Brown University},
 title = {On a Routing Problem},
 urldate = {2023-09-24},
 volume = {16},
 year = {1958}}

@book{Ford56,
author="Ford, L. R.",
title="Network Flow Theory.",
address="Santa Monica, CA",
year="1956",
doi="",
publisher="RAND Corporation"
}

@misc{BusinessRules23,
title = {Business Rules for Joint Seat Allocation for the Academic Programs offered by the IITs, NITs, IIEST, IIITs, and Other-GFTIs for the academic year 2023-24},
author = {{India Ministry of Education}},
year = {2023},
URL ={https://cdnbbsr.s3waas.gov.in/s313111c20aee51aeb480ecbd988cd8cc9/uploads/2023/06/2023061151.pdf},
}

@article{AM81,
  author={Aizerman, M. and Malishevski, A.},
  journal={IEEE Transactions on Automatic Control}, 
  title={General theory of best variants choice: Some aspects}, 
  year={1981},
  volume={26},
  number={5},
  pages={1030-1040},
  doi={10.1109/TAC.1981.1102777}
}

@article{KC82,
 ISSN = {00129682, 14680262},
 URL = {http://www.jstor.org/stable/1913392},
 author = {Alexander S. Kelso and Vincent P. Crawford},
 journal = {Econometrica},
 number = {6},
 pages = {1483--1504},
 publisher = {[Wiley, Econometric Society]},
 title = {Job Matching, Coalition Formation, and Gross Substitutes},
 urldate = {2025-01-20},
 volume = {50},
 year = {1982}
}

@article{CY17,
Author = {Chambers, Christopher P. and Yenmez, M. Bumin},
Title = {Choice and Matching},
Journal = {American Economic Journal: Microeconomics},
Volume = {9},
Number = {3},
Year = {2017},
Month = {August},
Pages = {126–47},
DOI = {10.1257/mic.20150236},
URL = {https://www.aeaweb.org/articles?id=10.1257/mic.20150236}}

@article{AS13,
Author = {Aygün, Orhan and Sönmez, Tayfun},
Title = {Matching with Contracts: Comment},
Journal = {American Economic Review},
Volume = {103},
Number = {5},
Year = {2013},
Month = {August},
Pages = {2050–51},
DOI = {10.1257/aer.103.5.2050},
URL = {https://www.aeaweb.org/articles?id=10.1257/aer.103.5.2050}}

@article{Blair88,
 ISSN = {0364765X, 15265471},
 URL = {http://www.jstor.org/stable/3689947},
 author = {Charles Blair},
 journal = {Mathematics of Operations Research},
 number = {4},
 pages = {619--628},
 publisher = {INFORMS},
 title = {The Lattice Structure of the Set of Stable Matchings with Multiple Partners},
 urldate = {2025-01-21},
 volume = {13},
 year = {1988}
}

@article{HM05,
Author = {Hatfield, John William and Milgrom, Paul R.},
Title = {Matching with Contracts},
Journal = {American Economic Review},
Volume = {95},
Number = {4},
Year = {2005},
Month = {September},
Pages = {913–935},
DOI = {10.1257/0002828054825466},
URL = {https://www.aeaweb.org/articles?id=10.1257/0002828054825466}}

@article{Kuhn55,
  author = {Harold W. Kuhn},
  title = {The Hungarian Method for the Assignment Problem},
  journal = {Naval Research Logistics Quarterly},
  year = {1955},
  volume = {2},
  pages = {83--97}
}

@article{Munkres57,
  author = {J. Munkres},
  title = {Algorithms for the Assignment and Transportation Problems},
  journal = {Journal of the Society for Industrial and Applied Mathematics},
  year = {1957},
  volume = {5},
  number = {1},
  pages = {32--38},
  month = {Mar}
}

@article{Many-to-many-KM,
  author = {Zhu, Haibin and Lou, Dongning and Zhang, Siqin and Zhu, Yu and Teng, Luyao and Teng, Shaohua},
  title = {Solving the Many to Many assignment problem by improving the Kuhn–Munkres algorithm with backtracking},
  journal = {Theoretical Computer Science},
  year = {2016},
  volume = {618},
  pages = {30-41},
}

@article{Many-to-One-KM,
  author={Zhu, Haibin and Zhou, MengChu and Alkins, Rob},
  journal={IEEE Transactions on Systems, Man, and Cybernetics - Part A: Systems and Humans}, 
  title={Group Role Assignment via a Kuhn–Munkres Algorithm-Based Solution}, 
  year={2012},
  volume={42},
  number={3},
  pages={739-750},
  doi={10.1109/TSMCA.2011.2170414}
}

\appendix

\section{Omitted Proofs}\label{appendix: proofs}

    \subsection{Proof of \cref{equivalence result}}\label{proof of equivalence result}
    
    Before giving the proof, we will state a useful result from matching theory.
    
    \begin{lemma}[Mendelsohn-Dulmage Theorem (\citeyear{Dulmage-Mendelsohn58})] \label{Mendelsohn-Dulmage Theorem}
    Let $X$ and $Y$ be two sides in a matching market. Let $\mu_1$ and $\mu_2$ be two matchings that comply with eligibility requirements. Then, there exists a third matching $\mu_3$ that complies with eligibility requirements such that $\mu_1(X) \subseteq \mu_3(X)$ and $\mu_2(Y) \subseteq \mu_3(Y)$.
    \end{lemma}
    
    The proof of \cref{equivalence result} is now immediate.
    
    \begin{proof}[Proof of \cref{equivalence result}]
    Consider the graph of patients and vaccines where there is an edge between a patient and a vaccine iff the patient is eligible for the category of the vaccine. Let $\mu''$ be a matching that is maximum in resource allocation. By \cref{Mendelsohn-Dulmage Theorem}, there exists a matching $\mu'$ that matches the set of patients who are matched under $\mu$, and the set of vaccines that are matched under $\mu''$, which would also be maximum in resource allocation by definition.
    \end{proof}
    
    \subsection{Proof of \cref{Max-in-Max correctness}}\label{proof of max-in-max correctness}
    We will start by giving some definitions we will use in the proof.
    \begin{definition}
    There are three different types of alternating chains in a given matching $\mu$:
    \begin{itemize}
        \item An alternating chain of even length is called a \textbf{neutral alternating chain} of $\mu$. 
        \item An alternating chain of odd length that starts and ends with an edge present in $\mu$ is called an \textbf{incremental alternating chain} of $\mu$.
        \item An alternating chain of odd length that starts and ends with an edge that is not present in $\mu$ is called a \textbf{decremental alternating chain} of $\mu$.
    \end{itemize}
    \end{definition}
    
    First, we prove the \emph{only if} $( \Rightarrow )$ part.
    
    \begin{lemma}[$ \Rightarrow $]\label{only if part of the correctness of max-in-max}
        If a matching is max-in-max, then it is maximum in resource allocation and there is no neutral alternating chain with positive potential. 
    \end{lemma}
    \begin{proof}
        Max-in-max implies maximum in resource allocation by definition. For the second part of the proof, we prove the contra-positive. Suppose that there is a neutral alternating chain 
         with positive potential in matching $\mu$. The matching $\mu'$ we obtain by augmenting this chain (1) matches a larger number of patients with a category they are beneficiaries of, and (2) has the same size as $\mu$. Therefore, the matching $\mu$ cannot be max-in-max, which concludes the proof. 
    \end{proof}
    
    We will prove the \emph{if} $ (\Leftarrow) $ part of the proof in the rest of this section. Let us start with a definition.
    \begin{definition}
         For any two given matchings $\mu_1$ and $\mu_2$, the \textbf{symmetric difference} of them is defined to be the set $$(\mu_1 - \mu_2) \cup (\mu_2 - \mu_1)$$ and it is denoted by $ \mu_1 \oplus \mu_2$.
    \end{definition}
    
    Next, we prove a lemma that generalizes an important decomposition result used in the study of augmenting paths.
    \begin{lemma}\label{connected components lemma}
        Let $\mu_1$ and $\mu_2$ be two arbitrary matchings. Then, $\mu_1 \oplus \mu_2$ can be decomposed into edge-disjoint subgraphs, each of which has one of the following forms: 
        \begin{enumerate}
            \item An incremental alternating chain of $\mu_1$ (equivalently, a decremental alternating chain of $\mu_2$), i.e. a path whose edges alternate between $\mu_1$ and $\mu_2$ and has more edges in $\mu_1$ than $\mu_2$.
            \item A decremental alternating chain of $\mu_1$ (equivalently, an incremental alternating chain of $\mu_2$), i.e. a path whose edges alternate between $\mu_1$ and $\mu_2$ and has more edges in $\mu_2$ than $\mu_1$.
            \item A neutral chain of $\mu_1$ (equivalently, a neutral chain of $\mu_2$), i.e. a path whose edges alternate between $\mu_1$ and $\mu_2$ and has equally many edges in $\mu_1$ and $\mu_2$.\footnote{An alternating cycle can be represented as a neutral chain that ends at the vertex it starts.}
        \end{enumerate}
    \end{lemma}
    \begin{proof}
        Note that a graph $G$ between patients and categories can be converted into a graph $G'$ between patients and vaccines in a canonical way by creating $r_c$ vaccines for each category $c$, and matching each patient with an arbitrary vaccine from the category she is matched to such that no vaccine is assigned to more than one patient.\footnote{Note that the new graph can be constructed in possibly more than one way, but is unique up to the labeling of vaccines. However, different labeling of vaccines would possibly result in different decompositions of the symmetric difference.} Let these new matchings be $\mu_1'$ and $\mu_2'$. The symmetric difference $\mu_1' \oplus \mu_2'$ is a graph with maximum degree less than or equal to two. If a vertex has degree two, one of its edges must be in $\mu_1'$ and the other one must be in $\mu_2'$, by construction. It is a well-known folklore result in graph theory that any graph where every vertex has degree at most two must consist of vertex-disjoint components of isolated vertices, paths, or cycles. Converting each component in $G'$ into a component in the original graph $G$ between patients and categories using a canonical transformation function that maps the vaccines to their respective categories yields a desired decomposition of the symmetric difference $\mu_1 \oplus \mu_2$.
    \end{proof}
    
    Now, we will continue with a lemma that is more specific to the domain of our result.
    \begin{lemma} \label{XOR of two maximal matchings}
        Let $\mu_1$ and $\mu_2$ be matchings that are maximum in resource allocation. Then every edge-disjoint component in a given decomposition of $\mu_1 \oplus \mu_2$ is a neutral chain. 
    \end{lemma}
    \begin{proof}
        Consider an arbitrary component. It must be in one of the forms stated in \cref{connected components lemma}. If it is an incremental alternating chain of $\mu_1$, then we can augment this chain in $\mu_1$ and obtain a new matching $\mu_1'$ that has size $\abs{\mu_1'} = \abs{\mu_1} + 1$, which contradicts with the assumption that $\mu_1$ is maximum in resource allocation. 
    
        Similarly, if it is an incremental alternating chain of $\mu_2$, then we can augment this chain in $\mu_2$ and obtain a new matching $\mu_2'$ that has size $\abs{\mu_2'} = \abs{\mu_2} + 1$, which contradicts with the assumption that $\mu_2$ is maximum in resource allocation. Therefore, every component must be either a neutral chain.
    \end{proof}

    Finally, we are ready to prove the \emph{if} $ (\Leftarrow) $ part, and conclude the proof.
    \begin{lemma}[$ \Leftarrow $]\label{if part of the correctness of max-in-max}
        If a matching $\mu$ is maximum in resource allocation and there is no neutral alternating chain with positive potential, then $\mu$ is also max-in-max.
    \end{lemma}
    \begin{proof}
     We prove the contra-positive. Namely, we prove that if a matching $\mu$ that is maximum in resource allocation but is not max-in-max, then there must be a neutral alternating chain with positive potential. 
    
     Let $\mu_{OPT}$ be a max-in-max matching. Consider an arbitrary decomposition of $\mu \oplus \mu_{OPT}$ in the form of \cref{connected components lemma}. Let the set of alternating chains in this decomposition be $S$. Then,
     \begin{align*}
     \sum_{s \in S} \Phi_{\mu}(s)
     &= \sum_{c \in \C} ( \bigm| \{i \in B_c \mid i \in \mu_{OPT}^{-1}(c) \} \bigm| -  \bigm| \{i \in B_c \mid i \in \mu^{-1}(c) \} \bigm| ) \\
     &= \sum_{c \in \C}  \bigm| \{i \in B_c \mid i \in \mu_{OPT}^{-1}(c) \} \bigm| -   \sum_{c \in \C} \bigm| \{i \in B_c \mid i \in \mu^{-1}(c) \} \bigm| \\
     &= b_{\mu_{OPT}} - b_{\mu}
     \end{align*}
    where $b_{\mu_{OPT}}$ and $b_{\mu}$ are the number of beneficiary assignments in $\mu_{OPT}$ and $\mu$, respectively. Here, the first equation follows by the fact that $S$ is an edge-disjoint decomposition of $\mu \oplus \mu_{OPT}$, the second follows by distributing the summation, and the third is by definition. This equation basically says that the sum of the potentials of alternating chains in the symmetric difference is equal to the difference between the number of beneficiaries who receive a vaccine under $\mu$ and the number of beneficiaries who receive a vaccine under $\mu_{OPT}$.
    
    Since $\mu_{OPT}$ is maximum in beneficiary assignment but $\mu$ is not, the final number we get must be positive, i.e.
    \begin{align*}
       \sum_{s \in S} \Phi_{\mu}(s) =  b_{\mu_{OPT}} - b_{\mu} > 0.
    \end{align*}
    Therefore, there must exist some $s^{+} \in S$ such that $\Phi_{\mu}(s^{+}) > 0$. However, $S$ consists of only neutral chains by \cref{XOR of two maximal matchings}. Hence, $s^{+}$ is a neutral alternating chain with positive potential in $\mu$.
    \end{proof}
    
    \begin{proof}[Proof of \cref{Max-in-Max correctness}]
    
    \cref{if part of the correctness of max-in-max} and \cref{only if part of the correctness of max-in-max} together conclude the proof of the correctness of Max-in-Max algorithm.
    \end{proof}

\subsection{Proof of \cref{thm: IMMAM properties}}\label{proof of IMMAM properties}

    Recall that we call the \emph{if} condition ($e' = e-1 \text{ and } b' = b \text{ or (}(p \in \B_c \text{ and } b' = b-1\text{)}$) in the description of IMMAM the \emph{permanent assignment condition.}
    \begin{proof}
    \emph{Universal domain.} IMMAM admits universal domain by construction since we do not impose any restrictions on the vaccine allocation market being input. 

    \emph{Compliance with eligibility requirements.} In the mechanism, the pairs $(p,c)$ such that $p$ is not eligible for category $c$ are never considered to be matched. Hence, by construction, IMMAM complies with eligibility requirements.

    \emph{Max-in-max.} The permanent assignment condition in IMMAM is designed to ensure that $(p,c)$ is matched if and only if there remains a max-in-max matching in the set of available matchings. Therefore, IMMAM satisfies max-in-max by construction.

    \emph{Respecting priorities.} Suppose the contrary. Let $p \: \pi_c \: p'$, but $\mu(p) = \varnothing$ and $\mu(p') = c$, where $\mu$ is the final outcome. Let $\mu'$ be the matching such that $\mu'(p) = c$, $\mu'(p') = \varnothing$, and $\mu'(p'') = \mu(p'')$ for all $p'' \in \PP \setminus \{p, p'\}$. Note that $\mu$ is max-in-max, and so $\mu'$. At the time IMMAM is processing the pair $(p,c)$, the matching $\mu'$ was available. Therefore, the permanent assignment condition would be satisfied, and IMMAM would permanently assign a vaccine to patient $p$ from category $c$, which leads to a contradiction.
\end{proof}
    
    \subsection{Proof of \cref{thm: path independence}}\label{section: proof of path independence}
    
    We start by giving two definitions, whose conjunction is equivalent to path independence.
    
    \begin{definition}[Consistency]\label{def: consistency}
        A choice rule $C(\cdot)$ is consistent if for every $X$ and $Y$ in the domain of $C(\cdot)$ such that $C(X) \subseteq Y \subseteq X$, we have $C(X) = C(Y)$.
    \end{definition}
    In the context of a vaccine allocation market, consistency implies that removing patients that are not assigned a vaccine does not change the outcome. It was first studied by \cite{Blair88}, which is also the first paper to study stable matchings with choice functions as the primitives.
    
    \begin{definition}[Substitutability]\label{def: substitutability}
        A choice rule $C(\cdot)$ is substitutable if for every $X$ and $Y$ in the domain of $C(\cdot)$ such that $Y \subseteq X$, we have $C(X) \cap Y \subseteq C(Y)$.
    \end{definition}
    
    Substitutability means that if a patient is assigned a vaccine now, they must still be assigned a vaccine if some other patients leave the market. It was first studied by \cite{KC82} in matching theory.
    
    \begin{lemma}[\cite{AM81}]\label{path independence = consistency and substitutability condition}
        A choice rule $C(\cdot)$ is path independent if and only if it is consistent and substitutable.
    \end{lemma}
    
    This equivalence result from the study of path independent choice functions helps us reduce the proof to showing that the resulting choice function $C^{\vartriangleright}_{(\PP, \C, r,\pi)}$ satisfies consistency and substitutability.\footnote{The fact that consistency and substitutability together are necessary and sufficient conditions for the stability of a mechanism was first shown by \cite{AS13}, who call consistency by the name of \emph{irrelevance of rejected contracts} in a matching with contracts \citep{HM05} setting. It was \cite{CY17} who pointed out and extensively studied the relationship between stability and path independence.}  We start by proving consistency. In the following and thereafter, we denote the outcome of IMMAM on a vaccine allocation market $(\PP, \CC, r, \pi)$ by $\M(\PP, \CC, r, \pi)$.\footnote{We exclude the input $\vartriangleright$ from the notation for simplicity since there is no room for confusion.}

        \begin{lemma}\label{lemma: excluding non-chosen patients yields the same outcome}
        Let $(\PP, \CC, r, \pi)$ be a vaccine allocation market, and $\PP' = C^{\vartriangleright}_{(\PP, \CC, r, \pi)}(\PP)$ be the set of patients who are chosen after running IMMAM. Then, for any set $\PP' \subseteq X \subseteq \PP$, we have $\M(\PP, \CC, r, \pi) = \M(X, \CC, r, \pi)$.
    \end{lemma}
    \begin{proof}
        We will show that for any patient $p \in \PP \setminus \PP'$, we have $\M(\PP, \CC, r, \pi) = \M(\PP \setminus \{p\}, \CC, r, \pi)$. First, we show that IMMAM makes the same decision for whether to match them or not for every pair $(p',c')$ prior to $(p,c)$ by using induction on the number of steps. The inductive step also proves the base step; thus, it suffices to prove the inductive step. Suppose IMMAM makes the same decision until it is time to process a pair $(p',c')$. We show that it also makes the same decision for this pair unless $(p',c') = (p,c)$.

        First, note that $\mu = \M(\PP, \CC, r, \pi)$ is also a feasible matching in $(\PP \setminus \{p\}, \CC, r, \pi)$ by definition of $p$. Moreover, any matching that is feasible in the smaller market is also feasible in the larger market. Therefore, the max-in-max numbers of these markets are the same. Thus, if there exists a max-in-max matching in $(\PP \setminus \{p\}, \CC, r, \pi)$ that respects the previously made assignments during IMMAM and matches the pair $(p',c')$, then IMMAM must also match $(p',c')$ since the same matching would also respect the previously made assignments in $(\PP, \CC, r, \pi)$ by induction, and is max-in-max.
        
        It remains to show that if IMMAM matches $(p',c')$ in the larger market, then it also does so in the smaller market. We know that the matching $\mu = \M(\PP, \CC, r, \pi)$ respects the previously made assignments by construction, and it does not match $p$ by definition. Therefore, $\mu$ is also available in $(\PP \setminus \{p\}, \CC, r, \pi)$ when IMMAM is processing $(p',c')$. Thus, if $(p',c') \in \mu$, IMMAM must also match $(p',c')$ in $(\PP \setminus \{p\}, \CC, r, \pi)$. This concludes the argument for the inductive step.

        Now, we know that the matching IMMAM has built until it is time to process $(p,c)$ in the larger market is the same in both markets. By definition of $p$, IMMAM will not match $(p,c)$. After processing $(p,c)$, the state of the two algorithms coincide. Thus, IMMAM must output the same matching in both markets. This concludes the proof that $\M(\PP, \CC, r, \pi) = \M(\PP \setminus \{p\}, \CC, r, \pi)$. 

        The proof of the more general statement given in the lemma immediately follows by induction on the number of patients being removed from the market.
    \end{proof}

    Note that the proof logic of this lemma holds even for excluding the \emph{pairs} that are not chosen instead of \emph{patients} that are not chosen. In fact, this result itself is stronger than the consistency of IMMAM. It says that not only the chosen set of patients are the same, but also the resulting matching is the same. We will use this lemma also in the proof of \cref{thm: outcome equivalence}. 

    \begin{corollary} \label{IMMAM is consistent}
        The choice function $C^{\vartriangleright}_{(\PP, \C, r, \pi)}(\cdot)$ that represents the chosen patients in the outcome of IMMAM is consistent.
    \end{corollary}

    Next, we prove a result that sheds light into the structure of the difference between the resulting matchings from two different sets of patients, \emph{ceteris paribus}.
    
    \begin{lemma} \label{single component lemma}
        Let $(\PP,\C, r,\pi)$ be a vaccine allocation market, and $X, Y \subseteq \PP$ be two sets of patients. Consider an arbitrary decomposition of the form in \cref{connected components lemma} of the symmetric difference $\M(X,\C, r,\pi) \oplus \M(Y,\C, r,\pi)$ of the matchings from the induced sub-markets. Every component in this decomposition must involve a patient that is not in $X \cap Y$.
    \end{lemma}
    \begin{proof}
        For ease of notation, let $\mu_X = \M(X,\C, r,\pi)$ and $\mu_Y = \M(Y,\C, r,\pi)$. Fix a decomposition of $\mu_X \oplus \mu_Y$. Suppose that there exists a component in this decomposition whose patient set is a subset of $X \cap Y$. Let $\mu_X'$ and $\mu_Y'$ be the resulting matchings after augmenting this component in matching $\mu_X$ and $\mu_Y$, respectively. We start by arguing that this component must be a neutral alternating chain with potential equal to 0, and then we finish the proof by using a well-ordering argument. 
        
        First, it must be a neutral alternating chain, since otherwise augmenting this chain in either $\mu_X$ or $\mu_Y$ would result in a matching that allocates more vaccines. Having noted this, it must also have potential equal to 0, since otherwise one could similarly augment this chain in either $\mu$ or $\mu'$ to obtain a matching that allocates the same number of vaccines but makes more beneficiary assignments. 
    
        Now, consider the patient-category pair $(p_0, c_0)$ in this component that has the highest priority with respect to $\succ$. Without loss of generality, assume that $(p_0, c_0) \in \mu_X$. Then, consider the point where we process $(p_0, c)_0$ when the mechanism $\M$ is run on the induced market $(Y,\C, r,\pi)$. Note that no patient-category pair that belongs to the component has yet been processed, so they are all still in the market. Thus, $\mu_Y'$ is still a matching that is permitted, and is max-in-max since this component is a neutral alternating chain with potential 0, as we argued above. Hence, $(p, c)$ must have been matched and left the market in this step. Therefore, $\mu_Y$ could not have been resulted by this mechanism, which concludes the proof.
    \end{proof}

        We can now prove that the substitutability condition holds.

        \begin{lemma} \label{IMMAM is substitutable}
            The choice function $C^{\vartriangleright}_{(\PP, \C, r, \pi)}$ that represents the chosen patients in the outcome of IMMAM is substitutable.
        \end{lemma}
        \begin{proof}
            Now, we will show that the mechanism satisfies the substitutes condition. By a straightforward inductive argument, it suffices to prove that $C^{\vartriangleright}_{(\PP, \C, r, \pi)}(X \cup \{y\}) \setminus \{y\} \subseteq C^{\vartriangleright}_{(\PP, \C, r, \pi)}(X)$ for any set of patients $X \subseteq \PP$ and $y \in \PP \setminus X$. Let $\mu = \M(X,\C, r,\pi)$, and $\mu' = \M(X \cup \{y\} ,\C, r,\pi)$. By \cref{single component lemma}, we have that the decomposition of $\mu \oplus \mu'$ must have a single alternating chain component and patient $y$ must be a part of this component. Note that $y$ is not matched under $\mu$ by definition. Therefore, $y$ must be one of the ends of the alternating chain, which means that all of the other patients in this alternating chain must be matched under both $\mu$ and $\mu'$. This implies that $C(X \cup \{y\}) \setminus \{y\} \subseteq C(X)$, as desired.
        \end{proof}
    
        We are now ready to finish the proof by putting together the tools we have developed.

        \begin{proof}[Proof of \cref{thm: path independence}]
        The proof follows from \cref{path independence = consistency and substitutability condition}, \cref{IMMAM is consistent}, and \cref{IMMAM is substitutable}.

        \end{proof}

    


    \subsection{Proof of \cref{thm: comparative statics}}\label{proof: comparative statics}
        The first half immediately follows from \cref{IMMAM is substitutable}. We will use the first half to prove the second by creating a hypothetical market and reducing the second statement to the first. We start by stating a tautological fact.
        
        \begin{fact}\label{equivalence of choices in induced markets}
            Let $(\PP, \C, r,\pi)$ and $(\PP', \C, r,\pi)$ be two vaccine allocation markets such that $\PP' \subseteq \PP$ and $\pi$ is defined over $\PP$. Then, for any $X \subseteq \PP'$, we have $C^{\vartriangleright}_{(\PP, \C, r,\pi)}(X) = C^{\vartriangleright}_{(\PP', \C, r,\pi)}(X)$.
        \end{fact}
        We will make use of this fact multiple times in the proof for manipulating the expressions. Next, we prove a lemma. Why this lemma might be of use will be more clear to the reader once we describe the hypothetical market in the proof.
        
        \begin{lemma}\label{lemma: relating the max-in-max pairs in two markets}
             Let $\pi$ be an arbitrary priority ordering defined on the disjoint union of patients $\PP \cup \{p\}$ for an arbitrary set of patients $\PP$ and a patient $p \not\in \PP$. Let $(\PP, \CC, r, \pi)$ and $(\PP \cup \{p\}, \CC, (r_{-c}, r_c + 1), \pi)$ be two vaccine allocation markets such that $p \not\in \PP$, $p \in \B_c$, and for all $c' \not\in c$, $p \not\in \E_c$, and let the max-in-max pair of $(\PP, \CC, r, \pi)$ be $(e, b)$. Then, the max-in-max pair of $(\PP \cup \{p\}, \CC, (r_{-c}, r_c + 1), \pi)$ is $(e+1, b+1)$.
        \end{lemma}
        \begin{proof}
            Let the max-in-max pair of $(\PP \cup \{{p}\}, \CC, (r_{-c}, r_c + 1), \pi)$ be $(e', b')$. First, note that $e \geq e' - 1$ and $b \geq b' - 1$ since for any matching $\mu'$ defined on the market $(\PP \cup \{{p}\}, \CC, (r_{-c}, r_c + 1), \pi)$, $\mu := \mu' \setminus \{({p}, \mu'(p))\}$ is a matching defined on the market $(\PP, \CC, r, \pi)$. However, it is also the case that for any matching $\mu$ in $(\PP, \CC, r, \pi)$, $\mu' := \mu \cup \{({p}, \mu'(p))\}$ is a matching in $(\PP \cup \{{p}\}, \CC, (r_{-c}, r_c + 1), \pi)$. Thus, $e' \geq e + 1$ and $b' \geq b + 1$.  Therefore, it must hold that $e' = e + 1$, $b' = b+1$.
        \end{proof}
    
        We stated the above fact and lemma prior to describing the hypothetical market and introducing more notation to make it clear that they do not only hold for the one instance at hand, but in general. Indeed, that will come handy in the proof.

        Now, we are ready to give a complete proof of \cref{thm: comparative statics}.
    
        \begin{proof}[Proof of \cref{thm: comparative statics}]
            As previously mentioned, the first statement follows from \cref{IMMAM is substitutable}. We prove the second statement in the theorem.
    
            Fix a category $c \in \C$. Now, we describe a hypothetical market $(\PP \cup \{\bar{p}\}, \CC, (r_{-c}, r_c + 1), \pi')$. Let $\bar{p}$ be a new patient, and $\pi'$ be a priority profile over the set of patients $\PP \cup \{\bar{p}\}$ such that for any $c' \in \C \setminus \{c\}$, we have $\E_{c'} \: \pi_{c'}' \bar{p}$, that for any $p \in \PP$ we have $\bar{p} \: \pi_c' \: p$ and $\bar{p} \: \pi_c' \: \beta_c$, and that $(\PP, \CC, r, \pi)$ is an induced sub-market of $(\PP \cup \{\bar{p}\}, \CC, (r_{-c}, r_c + 1), \pi')$. Note that these conditions define $\pi'$ uniquely up to the freedom allowed between $\bar{p}$ and the other patients below the eligibility threshold $\E_{c'}$ for any category $c' \in \C \setminus \{c\}$, and that this freedom is unimportant for the sake of the assignment process due to the eligibility condition.
            
            For any $X \subseteq \PP$, we have
            \begin{align*}
            C^{\vartriangleright}_{(\PP, \CC, (r_{-c}, r_c + 1), \pi)}(X) &= C^{\vartriangleright}_{(\PP \cup \{\bar{p}\}, \CC, (r_{-c}, r_c + 1), \pi')}(X) \\
            &\supseteq  C^{\vartriangleright}_{(\PP \cup \{\bar{p}\}, \CC, (r_{-c}, r_c + 1), \pi')}(X \cup \{\bar{p}\}) \setminus \{\bar{p}\} \\
            &= C^{\vartriangleright}_{(\PP \cup \{\bar{p}\}, \CC, r, \pi')}(X) \\
            &= C^{\vartriangleright}_{(\PP, \CC, r, \pi)}(X).
            \end{align*}
            Here, the first and the last equations follow from \cref{equivalence of choices in induced markets}. The second identity follows from the first half of this proposition (or \cref{IMMAM is substitutable}). It remains to show that for any $X \subseteq \PP$, we have
            $C^{\vartriangleright}_{(\PP \cup \{\bar{p}\}, \CC, (r_{-c}, r_c + 1), \pi')}(X \cup \{\bar{p}\}) \setminus \{\bar{p}\}
            = C^{\vartriangleright}_{(\PP \cup \{\bar{p}\}, \CC, r, \pi')}(X)$, and the rest of the proof is dedicated to that.

            The proof makes use of the dichotomy that $\bar{p}$ is only eligible for category $c$, and moreover it is a beneficiary thereof. We will prove the stronger statement that in fact IMMAM run on the market $(X \cup \{\bar{p}\}, \CC, (r_{-c}, r_c + 1), \pi')$ makes the same assignments to every single patient when it is run on the market $(X, \CC, r, \pi')$, except that it also assigns $\bar{p}$ to $c$.
            
            First, we will show by induction on the number of steps in IMMAM that the algorithm makes the same decision for whether to match a pair $(p',c')$ for each edge prior to $(\bar{p}, c)$ when it is run on market $(\PP \cup \{\bar{p}\}, \CC, (r_{-c}, r_c + 1), \pi')$ and $(\PP, \CC, r, \pi')$ with respect to the precedence relation $\succ$. Our inductive step also proves the base step, therefore there is no need to prove the base step separately. Suppose the hypothesis holds up until when the algorithm is processing a pair $(p', c')$. Now, consider the two induced markets after removing the patients and vaccines who are matched from both of the markets. By the induction hypothesis, the new markets must be in the form $(\PP' \cup \{\bar{p}\}, \CC, (r_{-c}', r_c' + 1), \pi')$ and $(\PP', \CC, r', \pi')$ for some $\PP' \subseteq \PP$ and $r' \leq r$ where the inequality is point-wise. We will show that there is a max-in-max matching that contains $(p', c')$ in the smaller market if and only if there exists such a matching in the larger market. Suppose there is a max-in-max matching $\mu$ in $(\PP', \CC, r', \pi')$ that contains the pair $(p', c')$. Then, the matching $\mu'$ such that $\mu'(p) = \mu(p)$ for all $p \neq \bar{p}$, and $\mu'(\bar{p}) = c$ is also a max-in-max matching in $(\PP' \cup \{\bar{p}\}, \CC, (r_{-c}', r_c' + 1), \pi')$ by \cref{lemma: relating the max-in-max pairs in two markets}. This concludes the \emph{only if} part of the argument. Next, suppose that there is a max-in-max matching $\mu$ in $(\PP' \cup \{\bar{p}\}, \CC, (r_{-c}', r_c' + 1), \pi')$ that contains the pair $(p', c')$. If $\mu(\bar{p}) = c$, then the matching $\mu'$ such that $\mu'(p) = \mu(p)$ for all $p \neq \bar{p}$ is max-in-max in the market $(\PP', \CC, r', \pi')$ by \cref{lemma: relating the max-in-max pairs in two markets}. If $\mu(\bar{p}) = \varnothing$, we know that $\mu(c)$ is non-empty since we could otherwise match $\bar{p}$, which would contradict with the assumption that $\mu$ is max-in-max (in fact, it must be that $\mu$ assigns all of the vaccines in $c$ by the same argument). Then, let $p'' \in \mu(c)$ be an arbitrary patient, and define $\mu'$ such that $\mu'(p) = \mu(p)$ for all $p \neq p''$, and $\mu'(p'') = \varnothing$. By \cref{lemma: relating the max-in-max pairs in two markets}, $\mu''$ is a max-in-max matching in $(\PP', \CC, r', \pi')$. This concludes the proof that the algorithm makes the same decision for every pair $(p',c')$ until it is time to process the pair $(\bar{p}, c)$.
    
            When it is time to process the pair $(\bar{p}, c)$, the algorithm being run on market $(\PP \cup \{\bar{p}\}, \CC, (r_{-c}, r_c + 1), \pi')$ must match $\bar{p}$ to $c$. For that, we will show that there always exists a max-in-max matching that respects the previously made assignments in the algorithm and matches $\bar{p}$ to $c$. Note that $\bar{p}$ has not been matched yet since it is only eligible for category $c$, and no vaccine from category $c$ has been matched yet since $\bar{p}$ is the first patient to be processed in $c$. Let $\mu$ be an arbitrary max-in-max matching that respects the previously made assignments. If $\mu(\bar{p}) = c$, there is nothing to show. If $\mu(\bar{p}) = \varnothing$, then there must be a patient $p' \neq \bar{p}$ such that $\mu(p)$ since otherwise we could match $\bar{p}$ to $c$. Then, the matching $\mu'$ such that $\mu'(p) = \mu(p)$ for all $p \neq \bar{p}, p'$, $\mu'(\bar{p}) = p$, and $\mu'(p') = \varnothing$ respects the previously made assignments, matches $\bar{p}$ to $c$, and is max-in-max, as desired.
    
            Observe that after the pair $(\bar{p}, c)$ has been processed, the states of the two markets coincide with each other. Since the algorithm is deterministic, the rest of the algorithm must assign the same patient-category pairs, which concludes the proof.
        \end{proof}

    \subsection{Proof of \cref{thm: outcome equivalence}} \label{proof: outcome equivalence}

    The following proof assumes that IMMAM and IMMAM-M are outcome-equivalent. Since IMMAM-M consists of a simpler optimization that leaves the general structure of IMMAM mostly the same, we do not prove it as a separate result; nonetheless, we briefly justify it in \cref{appendix: IMMAM-M pseudo}.
    
    \begin{proof}
        We prove it by induction on $\ceil{\log_2 (|\PP|/q)}$. The result is a tautology when $\ceil{\log_2 (|\PP|/q)} \leq 1$. Now, we prove that if it is true for any $\PP$ such that $\ceil{\log_2 (|\PP|/q)} = k$ for some integer $k \geq 1$, then it is also true for any $\PP$ such that $\ceil{\log_2 (|\PP|/q)} = k+1$. Note that $\PP_1$ and $\PP_2$ we obtain by splitting $\PP$ into two sets of roughly equal size satisfy $\ceil{\log_2 (|\PP_1|/q)} = \ceil{\log_2 (|\PP_2|/q)} = k$. Therefore, by induction hypothesis, $\PP_1^{\text{(chosen)}} = C^{\vartriangleright}_{(\PP,\C, r,\pi)}(\PP_1)$, and $\PP_2^{\text{(chosen)}} =  C^{\vartriangleright}_{(\PP,\C, r,\pi)}(\PP_2)$. By path independence (\cref{thm: path independence}), we have $C^{\vartriangleright}_{(\PP,\C, r,\pi)}(\PP) = C^{\vartriangleright}_{(\PP,\C, r,\pi)}(C^{\vartriangleright}_{(\PP,\C, r,\pi)}(\PP_1) \cup C^{\vartriangleright}_{(\PP,\C, r,\pi)}(\PP_2) = C^{\vartriangleright}_{(\PP,\C, r,\pi)}(\PP_1^{\text{(chosen)}} \cup \PP_2^{\text{(chosen)}})$. In particular, this implies that $C^{\vartriangleright}_{(\PP,\C, r,\pi)}(\PP) \subseteq \PP_1^{\text{(chosen)}} \cup \PP_2^{\text{(chosen)}}$. Thus, by \cref{lemma: excluding non-chosen patients yields the same outcome}, running IMMAM-M (or IMMAM) on the induced sub-market $(\PP_1^{\text{(chosen)}} \cup \PP_2^{\text{(chosen)}}, \C, r,\pi)$ yields the same outcome as running it on $(\PP,\C, r,\pi)$.
    \end{proof}

\section{Omitted Algorithms} \label{appendix: omitted algorithms}
    \subsection{Algorithm for Finding an Even Alternating Chain with Positive Potential} \label{appendix: finding augmenting path in poly time}

In the description of the Max-in-Max algorithm as given in \cref{Max-in-Max Subsection}, it is not immediately clear whether we can find an even alternating chain with positive potential in a computationally efficient way. In this section, we briefly present an algorithm to find such a chain (if there is one) in polynomial time. The idea is to reduce the problem to finding a path with the minimum weight in a directed graph, and then use the Bellman-Ford algorithm \citep{Bellman58, Ford56}. Given a directed graph with possibly negative weights and a source vertex $s$, Bellman-Ford gives a path with minimum weight between $s$ and any other vertex. If there is a negative cycle, which implies that the minimum weight path is not well-defined, Bellman-Ford algorithm detects it and returns the cycle.

We will construct two different graphs for category-initiated alternating chains and patient-initiated alternating chains. First, we describe a directed graph in which finding a negative path corresponds to finding a category-initiated alternating chain.

For any given maximum matching $\mu$, we construct the following directed graph $G_{\mu}^{(1)} = (V^{(1)},E^{(1)})$. We start by describing the vertex set $V^{(1)}$.

\begin{itemize}
    \item For every category $c$, there is a vertex $v_c^{(1)}$.
    \item For every patient $p$, there is a vertex $v_p^{(1)}$.
    \item There is a source vertex $s^{(1)}$.
    \item There is a target vertex $t^{(1)}$.
\end{itemize}

The reason we have the source vertex is that when we construct the graph, we will use it as a \emph{master vertex} since Bellman-Ford is a single-source shortest path algorithm, and we use the target vertex to ensure that the outcome is an even (or neutral) chain. Next, we describe the edges and the weights in this graph.

\begin{itemize}
    \item For every category $c$ that has at least one unassigned vaccine under $\mu$, there is a directed edge of weight $0$ from the source vertex $s^{(1)}$ to $v_c^{(1)}$.
    \item For every category $c$, there is a directed edge of weight $0$ from $v_c^{(1)}$ to the target vertex $t^{(1)}$.
    \item For every category $c$ and patient $p$, there is a directed edge of weight $0$ from $v_c^{(1)}$ to $v_p^{(1)}$ if $p \in \E_c - \B_c$ but $p \not \in \mu(c)$.
    \item For every category $c$ and patient $p$, there is a directed edge of weight $-1$ from $v_c^{(1)}$ to $v_p^{(1)}$ if $p \in \B_c$ but $p \not \in \mu(c)$.
    \item For every patient $p$ that is matched under $\mu$, there is a directed edge of weight $0$ from $v_p^{(1)}$ to $v_{\mu(p)}^{(1)}$ if $p \not \in \B_c$.
    \item For every patient $p$ that is matched under $\mu$, there is a directed edge of weight $1$ from $v_p^{(1)}$ to $v_{\mu(p)}^{(1)}$ if $p \in \B_c$.  
\end{itemize}

Analogously, we define $G_{\mu}^{(2)} = (V^{(2)},E^{(2)})$ for any given maximum matching $\mu$ such that finding a negative path corresponds to finding a patient-initiated alternating chain. The graph construction is the same, except that we relabel patients as categories, and categories as patients. However, to avoid confusion, we will explicitly give the construction. Again, we start with the vertex set $V^{(2)}$.

\begin{itemize}
    \item For every patient $p$, there is a vertex $v_p^{(2)}$.
    \item For every category $c$, there is a vertex $v_c^{(2)}$.
    \item There is a source vertex $s^{(2)}$.
    \item There is a target vertex $t^{(2)}$.
\end{itemize}

Next, we describe the edge set $E^{(2)}$ in the same manner.

\begin{itemize}
    \item For every patient $p$ that is unassigned under $\mu$, there is a directed edge of weight $0$ from the source vertex $s^{(2)}$ to $v_p^{(2)}$.
    \item For every patient $p$, there is a directed edge of weight $0$ from $v_p^{(2)}$ to the target vertex $t^{(2)}$.
    \item For every patient $p$ and category $c$, there is a directed edge of weight $0$ from $v_p^{(2)}$ to $v_c^{(2)}$ if $p \in \E_c - \B_c$ but $p \not \in \mu(c)$.
    \item For every patient $p$ and category $c$, there is a directed edge of weight $-1$ from $v_p^{(2)}$ to $v_c^{(2)}$ if $p \in \B_c$ but $p \not \in \mu(c)$.
    \item For every category $c$ and patient $p \in \mu(c)$, there is a directed edge of weight $0$ from $v_c^{(2)}$ to $v_{p}^{(2)}$ if $p \not \in \B_c$.
    \item For every category $c$ and patient $p \in \mu(c)$, there is a directed edge of weight $1$ from $v_c^{(2)}$ to $v_{p}^{(2)}$ if $p \in \B_c$.  
\end{itemize}

Now, let us formally describe our algorithm. The following algorithm can be applied to $G_{\mu}^{(1)}$ and $G_{\mu}^{(2)}$ separately to find whether there is any category-initiated and patient-initiated even alternating chain with positive potential. Thus, in the algorithm below, we abuse the notation and drop the superscripts.




\begin{algorithm}[H]
\caption{Finding an Even Alternating Chain with Positive Potential}
\KwInput{A maximum matching $\mu$}
\KwOutput{An even alternating chain or a cycle with positive potential}
Construct a weighted directed graph $G_{\mu}$ as described above;\\
Run Bellman-Ford from the source vertex $s$. \\
\If{a negative cycle is detected, }{\KwRet that cycle.}
\Else{
\If{the shortest path from $s$ to $t$ has negative weight, }{\KwRet that path}
\Else{ \KwRet \textbf{False}}
}
\end{algorithm}

There is a one-to-one correspondence between cycles or paths from $s^{(1)}$ to $t^{(1)}$ (or $s^{(2)}$ to $t^{(2)}$) with negative weights in $G_{\mu}^{(1)}$ (or $G_{\mu}^{(2)}$), and alternating cycles or category-initiated (or patient-initiated) even alternating chains with positive potential in $\mu$. If there is a negative cycle present in the graph, Bellman-Ford algorithm detects it. Otherwise, it returns the shortest path. If the shortest path has negative weight, then we found an alternating chain with positive potential. If the shortest path has non-negative weight, then it means that there is no alternative chain with positive potential in this graph. This algorithm runs in $O(|\PP|\cdot|\C|^2)$ since the number of edges is $O(|\PP|\cdot|\C|)$ and the maximum length of any shortest path is $O(|C|)$ as the constructed graphs are bi-partite.

\subsection{IMMAM with Memory (IMMAM-M)}\label{appendix: IMMAM-M pseudo}

Several notes of clarification are in order with respect to \cref{caching mechanism}. First, when we augment the chain $(\mu(p), p, c, \mu^{-1}(c))$ in line 13, the last argument $\mu^{-1}(c)$ corresponds to a set. If $c$ has an unassigned vaccine, we consider $\mu^{-1}(c)$ to be a null patient, and thus the chain has length two;\footnote{The term \say{length} refers to the number of edges in the chain.} otherwise, we pick an arbitrary patient. Moreover, it is also possible that $\mu(p) = \varnothing$, in which case the chain has length two, again. However, for the sake of succinctness, we skip this case analysis and simply write \say{Augment the chain $(\mu(p), p, c, \mu^{-1}(c))$.}

The second note is concerning applying Max-in-Max on $\mu'$. In the Max-in-Max algorithm, we start with a maximum matching, and iteratively augment neutral alternating chains with positive potential until there are none remaining. Note that $\mu'$ in \cref{caching mechanism} before we apply the Max-in-Max algorithm is indeed a maximum matching of the induced sub-market at each iteration.

Finally, we briefly expand on the algorithm, through which we justify both its correctness and its runtime improvement. First, the algorithm matches \( p \) to \( c \) only if this assignment is part of a max-in-max matching that respects previously made assignments, and therefore the outcome is identical to that of the original IMMAM. Second, in cases where augmenting the chain \( (\mu(p), p, c, \mu^{-1}(c)) \) in line 13 results in the matching \( \mu' \) having one fewer assignment, lines 15--16 are used to recover another assignment. However, this augmentation may cause the matching to lose some of its beneficiary assignments---namely, at most half the length of the augmented chain, which is at most \( |\CC| \). Therefore, when the algorithm runs Max-in-Max in line 17, it needs to call the augmenting path finder at most \( |\CC| \) times. This is the key improvement in the algorithm's runtime.

\begin{algorithm}[H]
\caption{Iterative Max-in-Max Assignment Mechanism with Memory} \label{caching mechanism}
\KwInput{a vaccine allocation market $(\PP_0,\C_0, r_0,\pi_0)$, and a precedence list $\vartriangleright$ of categories}
\KwOutput{a matching $\mu$}
Using the Max-in-Max algorithm, find a max-in-max matching $\mu_0$ and the max-in-max pair $(e_0, b_0)$ of the vaccine allocation market $(\PP_0,\C_0, r_0,\pi_0)$;\\
Start with the max-in-max matching $\mu \leftarrow \mu_0$, the vaccine allocation market $(\PP,\C, r,\pi) \leftarrow (\PP_0,\C_0, r_0,\pi_0)$, and the max-in-max pair $(e, b) \leftarrow (e_0, b_0)$; \\
\For{$(p, c)$ in $\{(p, c) \mid p \in \E_{c}\}$ with the associated order of iteration $\succ_{(\PP_0, \CC_0, \pi_0, \vartriangleright)}$}{
\If{\emph{$p \not \in \PP$ \textbf{or} $r_c = 0$}}
{Skip to the next pair.}
\If{$\mu(p) = c$}
{
        Assign $\mu \leftarrow \mu \cup \{(p, c)\}$; \\
        Assign $e \leftarrow e - 1$; \\
        \If{$p \in \B_c$}
        {
        Assign $b \leftarrow b - 1$; \\
        }
        Assign $(\PP,\C, r,\pi) \leftarrow (\PP \setminus \{p\},\C, (r_{-c}, r_c - 1),\pi)$;
}

\If{$\mu(p) \neq c$}{
Augment the chain $(\mu(p), p, c, \mu^{-1}(c))$ in $\mu$ to obtain a new matching $\mu'$; \\
Tentatively decrease $r_c$ by one, and remove patient $p$ from the market to obtain an induced vaccine allocation market $(\PP \setminus \{p\},\C, (r_{-c}, r_c - 1),\pi)$; \\
\If{there is an incremental alternating chain $s$ in $\mu'$}{
    Augment $s$ in $\mu'$; \\
    }
    Starting from $\mu'$, apply Max-in-Max algorithm on $\mu'$ and let the max-in-max pair be $(e', b')$; \\
    \If{\emph{$e' = e-1$ \textbf{and} ($b' = b$ \textbf{or} ($p \in \B_c$ \textbf{and} $b' = b-1$))}}
    {
        Assign $\mu \leftarrow \mu \cup \{(p, c)\}$; \\
        Assign $e \leftarrow e'$ and $b \leftarrow b'$; \\
        Assign $(\PP,\C, r,\pi) \leftarrow (\PP \setminus \{p\},\C, (r_{-c}, r_c - 1),\pi)$;
    }
}

}

\KwRet $\mu$;
\end{algorithm}

\end{document}